\documentclass[superscriptaddress,twocolumn, amsmath, amssymb, aps]{revtex4-2}

\usepackage[dvipsnames]{xcolor}
\usepackage{graphicx}
\usepackage{enumitem}
\usepackage{siunitx}
\usepackage{booktabs}
\usepackage{multirow}
\usepackage{mathtools}
\usepackage[switch]{lineno}
\usepackage{comment}
\newcommand\myshade{30}
\colorlet{mylinkcolor}{red}
\colorlet{mycitecolor}{orange}
\colorlet{myurlcolor}{orange}

\usepackage[%
bookmarks=true,
colorlinks,
linkcolor=mylinkcolor!\myshade!black,
urlcolor=myurlcolor!\myshade!black,
citecolor=mycitecolor!\myshade!black,
plainpages=false,
pdfpagelabels,
final,
breaklinks=true
]{hyperref}

\usepackage[normalem]{ulem}
\newcommand{\clsout}[1]{\textcolor{red!80!black}{\sout{}}}

\usepackage{physics}
\usepackage{braket}

\usepackage[T1]{fontenc}
\usepackage{xspace}
\usepackage{soul}
\usepackage{float}
\begin{document}
\title{
Experimental observation of Higgs oscillations and their power-law decay across pairing regimes in a cavity QED quantum simulator}

\author{Chitose Maruko}
\thanks{These authors contributed equally to this work.}
\affiliation{JILA, NIST and University of Colorado, Boulder, Colorado 80309, USA}
\affiliation{Department of Physics, University of Colorado, Boulder, Colorado 80309, USA}

\author{Haoqing Zhang}
\thanks{These authors contributed equally to this work.}
\affiliation{JILA, NIST and University of Colorado, Boulder, Colorado 80309, USA}
\affiliation{Department of Physics, University of Colorado, Boulder, Colorado 80309, USA}

\author{Eliot A.~Bohr}
\thanks{These authors contributed equally to this work.}
\affiliation{JILA, NIST and University of Colorado, Boulder, Colorado 80309, USA}
\affiliation{Department of Physics, University of Colorado, Boulder, Colorado 80309, USA}
\thanks{These authors contributed equally to this work.}

\author{Leah Huzjak}
\affiliation{JILA, NIST and University of Colorado, Boulder, Colorado 80309, USA}
\affiliation{Department of Physics, University of Colorado, Boulder, Colorado 80309, USA}

\author{Anjun Chu}
\affiliation{Pritzker School of Molecular Engineering, University of Chicago, Chicago, Illinois 60637, USA}

\author{Victor Gurarie}
\affiliation{Department of Physics, University of Colorado, Boulder, Colorado 80309, USA}
\affiliation{Center for Theory of Quantum Matter, University of Colorado, Boulder, Colorado 80309, USA}

\author{Chengyi Luo}
\affiliation{Thomas J.~Watson, Sr., Laboratory of Applied Physics,  Kavli Nanoscience Institute, and Institute for Quantum Information and Matter,
California Institute of Technology, Pasadena, CA 91125, USA}

\author{Ana Maria Rey}
\affiliation{JILA, NIST and University of Colorado, Boulder, Colorado 80309, USA}
\affiliation{Department of Physics, University of Colorado, Boulder, Colorado 80309, USA}
\affiliation{Center for Theory of Quantum Matter, University of Colorado, Boulder, Colorado 80309, USA}

\author{James K.~Thompson}
\email{jkt@jila.colorado.edu}
\affiliation{JILA, NIST and University of Colorado, Boulder, Colorado 80309, USA}
\affiliation{Department of Physics, University of Colorado, Boulder, Colorado 80309, USA}

\date{\today}

\begin{abstract}
Understanding how pairing correlations in superconductors evolve out of equilibrium is a central challenge in quantum many-body physics. In systems with $s$-wave pairing, the character of Cooper pairing evolves continuously from a BCS regime of weakly bound, overlapping pairs to a BEC regime of tightly bound molecules, each with distinct excitation spectra. Here we realize and probe pairing dynamics across these regimes using cold, non-degenerate $^{87}$Rb atoms coupled to an optical cavity with Floquet dressing. Our cavity-QED platform emulates fermionic superfluids through a mapping onto Anderson pseudospins, while optical and microwave control tunes the ratio of interactions to single-particle energies, providing a flexible approach complementary to ultracold Feshbach and ultrafast solid-state platforms. Following a quench, we observe persistent pairing and transient oscillations of the order parameter associated with a collective amplitude (Higgs) mode. We find that the decay of these oscillations sharply distinguishes weak- and strong-pairing regimes: in the weak-pairing regime, oscillations decay with a power-law exponent $|\alpha|=0.65(8)$, whereas in the strong-pairing regime the decay is substantially faster, with $|\alpha|>1$. The oscillation frequency is set by the minimum quasiparticle excitation energy, directly linking nonequilibrium dynamics to the underlying excitation spectrum.
Our results establish nonequilibrium dynamics as a direct probe of pairing character and collective excitations in  interacting quantum systems.
\end{abstract}

\maketitle
\section{Introduction}

Collective phenomena arising from fermionic pairing underlie some of the most remarkable states of matter, including superconductors, superfluids and excitonic condensates. Within Bardeen--Cooper-Schrieffer (BCS) theory, attractive interactions produce a broken-symmetry state characterized by a complex order parameter or pairing energy gap  $\hbar \Delta$, whose magnitude sets the energy scale that protects the paired state. When driven out of equilibrium, this order parameter can exhibit rich collective dynamics, including three distinct dynamical phases, one of which is characterized by transient  oscillations of the amplitude of the order parameter,$|\Delta(t)|$,  the so-called Higgs mode \cite{Bardeen1957,Yuzbashyan2006,Barankov2006,Yuzbashyan2006Dynamical,Gurarie2009,Foster2013,Yuzbashyan2015}, which is  analogous to the Higgs boson in particle physics \cite{Higgs1964,Littlewood1982,Volkov1974}.
Although the three dynamical phases predicted by theory have recently been observed in a cavity-based quantum simulator \cite{Thompson2023BCS}, and Higgs oscillations have been detected in fermionic superfluids \cite{behrle2018higgs,kell2024exciting,Vale2024}, directly probing the intrinsic non-equilibrium dynamics of the Higgs mode has remained experimentally elusive. 
In particular, the Higgs mode is expected to encode universal signatures of the underlying  microscopic structure of the paired state, yet these features are difficult to resolve experimentally because the mode resides within the pair-breaking continuum and couples only weakly to conventional probes.

\begin{figure}[!ht]
\centering
\includegraphics[width=0.35\textwidth]{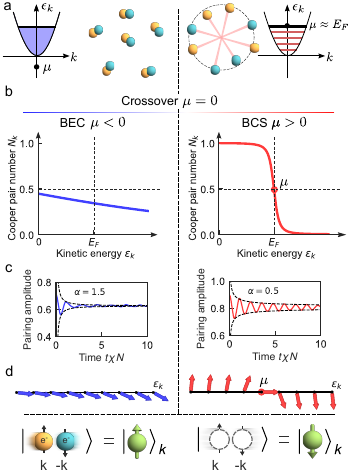}
\caption{\textbf{The BEC--BCS crossover, Higgs dynamics, and Anderson pseudospin mapping.}\small{
\textbf{a.} Schematic illustration of the BEC--BCS crossover obtained by tuning the attractive interaction between fermions. Strong attraction produces tightly bound molecular pairs that undergo Bose--Einstein condensation (BEC), whereas weak attraction leads to the formation of extended Cooper pairs and a Bardeen--Cooper--Schrieffer (BCS) superfluid. Across the crossover, the chemical potential changes sign from $\mu<0$ in the BEC regime to $\mu>0$ in the BCS regime.
\textbf{b.} Cooper-pair occupation $N_k$ as a function of kinetic energy $\epsilon_k$.  In the BCS regime, the pairing happens near the remnant Fermi level at the $\epsilon_k$ such $N_k=1/2$. In the BEC regime, $N_k<1/2$ corresponding to a negative chemical potential
\textbf{c.} Following a sudden change in interaction strength, the magnitude of the pairing energy gap $|\Delta(t)|$ exhibits coherent amplitude (Higgs) oscillations about its long-time value $\Delta_\infty$. These collisionless oscillations decay algebraically as $|\Delta(t)-\Delta_\infty|\propto t^{-\alpha}$, with the universal exponent $\alpha=1/2$ in the BCS regime. Across the crossover, the oscillations become increasingly damped and the exponent grows continuously, for example, approaching $\alpha=3/2$ in the deep BEC limit for a 3D system.
\textbf{d.} Anderson pseudospin representation of the reduced BCS model. For each momentum pair $(\mathbf{k},-\mathbf{k})$, the pseudospin-up state corresponds to an occupied Cooper-pair state and the pseudospin-down state to an empty state. Modes in a coherent superposition of occupied and empty (e.g. a nonzero pairing amplitude for such modes) are represented by spin or Bloch vectors with a nonzero transverse projection.  
}
}
\label{fig:fig1_theory}
\end{figure}
Importantly, for isotropic $s$-wave  pairing the decay of the Higgs oscillations is predicted to change  as weakly bound Cooper pairs (BCS) continuously evolve into a Bose-Einstein condensate (BEC) of tightly bound molecules (Fig.~\ref{fig:fig1_theory}a and b) \cite{greiner2003emergence,jochim2003bose,zwierlein2004condensation,regal2004observation}. In the BCS regime,  the Higgs oscillations are  predicted to exhibit a universal power-law  decay  exponent $\alpha=1/2$ , i.e  their amplitude damps with time as  $t^{-\alpha}$,  while in the BEC regime the oscillation decay exponent  is predicted to be  larger, i.e $t^{-\alpha}$  with $\alpha> 1/2$ (Fig.~\ref{fig:fig1_theory}c)  \cite{Gurarie2009}. Despite these clear predictions, the expected variation and exponents have remained experimentally elusive, largely due to density inhomogeneity, finite temperature and pair-breaking processes in ultracold Fermi gases \cite{Vale2024}.

Here we realize an effective $s$-wave BCS Hamiltonian in a cavity-mediated quantum simulator using the Anderson pseudospin mapping, which encodes Cooper pairs and holes into an effective spin-$1/2$ degree of freedom (See Fig.~\ref{fig:fig1_theory}d and \ref{fig:fig2_Anderson}a).  In contrast to prior implementations with atoms in cavities, where the power-law decay of the Higgs oscillations was obscured by the inhomogeneous atom-cavity couplings arising from the atoms being trapped in an incommensurate 1D optical lattice \cite{Thompson2023BCS, Young2025}, here we achieve both spatially homogeneous atom-cavity coupling and prevent undesirable motional sideband transitions  by allowing the atoms to fall freely along the cavity axis. Following a controlled quench, with the help of  continuous Floquet engineering (Fig.~\ref{fig:fig2_Anderson}b), we observe the stabilization of the pairing gap (Fig.~\ref{fig:fig2_Anderson}d) together with coherent Higgs oscillations (Fig.~\ref{fig:fig3}d) with an amplitude that  decays following   a power-law $t^{-\alpha}$, and a frequency set by the excitation spectra  following  theory predictions (Fig.~\ref{fig:fig3}b,c).  By tuning the effective interaction regime, we measure exponents evolving from $\alpha = 0.65(8)$ in the BCS regime to $\alpha \gtrsim 1.5$ in the BEC regime, in close agreement with theoretical calculations without free parameters (Figs.~\ref{fig:fig4}d and  \ref{fig:fig5}a). Our results establish the long-sought observation of universal Higgs-mode decay dynamics across the BCS--BEC crossover and demonstrate the power of cavity-QED quantum simulators for probing non-equilibrium collective phenomena.

\section{ The BEC-BCS crossover in the Anderson Pseudo-spin Mapping}
We directly connect our experimental spin system with spin-exchange interactions to the one-channel BCS Hamiltonian using the Anderson pseudo-spin mapping \cite{Anderson1958}.  The presence (or absence) of a Cooper pair formed by a pair of opposite spin electrons with momenta $\pm \mathbf{k}$ is mapped to a pseudospin in up (or down) for some mode indexed by $\mathbf{k}$, visualized as Bloch vectors for each mode in Fig.~\ref{fig:fig1_theory}d. At the very heart of superconductivity is that the system lowers its total energy by the modes being in quantum superpositions of both occupied and unoccupied by a Cooper pair, or a Bloch vector for the pseudospin that is tilted away from vertical.

The mapping   relates the annihilation operator of a Cooper pair  to a spin lowering operator by the relation $\hat{s}_{-,\mathbf{k}} =\hat{c}_{\mathbf{k},\sigma} \hat{c}_{\mathbf{-k},\bar{\sigma}}$ where \(\hat c_{\mathbf{k},\sigma}\) annihilates a fermion with momentum
\(\mathbf{k}\) and spin \(\{\sigma= \mathrm{up}, \bar{\sigma}=\mathrm{down}\}\). The average  Cooper-pair occupation  $N_\textbf{k}= \langle   \hat  N_\textbf{k} \rangle= \langle\hat c_{\mathbf{k},\sigma}^\dagger
\hat c_{\mathbf{k},\sigma}+ \hat c_{\mathbf{-k},\bar{\sigma}}^\dagger
\hat c_{-\mathbf{k},\bar{\sigma}}\rangle/2$ is linked to the pseudospin projection operator $\hat{s}_{z,\mathbf{k}} = \hat{N}_\mathbf{k} -1/2$, with eigenvalue  $+1/2$ if  there  is a Cooper pair  at momentum $(\mathbf{k,-k})$,  and $-1/2$  if the state is empty.  We also define collective spin operators $\hat{S}_\beta= \sum_{\textbf{k}=1}^N \hat{s}_{\beta,\mathbf{k}}$ with $\beta\in\left\{x, y, z, +, - \right\}$, where we have truncated the number of contributing modes to $N$.  The single mode $\hat{\mathbf{s}}_\mathbf{k}=\{\hat{s}_{x,\mathbf{k}},\hat{s}_{y,\mathbf{k}},\hat{s}_{z,\mathbf{k}}\}$ and collective $\hat{\mathbf{S}}=\{\hat{S}_x,\hat{S}_y,\hat{S}_z\}$ spin operators determine the single-mode $\langle \hat{\mathbf{s}}_{\mathbf{k}}\rangle$ and collective $\langle \hat{\mathbf{S}}\rangle$ Bloch vectors.

In terms of Anderson pseudo-spins, the one channel BCS model maps onto an all-to-all spin-exchange Hamiltonian 
\begin{equation}\label{eq:hamiltonianBCS}
\hat{H}_{\rm 1chan}/\hbar = 2\sum_{\bf k} \epsilon_{\mathbf{k}} \hat{s}_{z,\mathbf{k}} - \chi_F \hat{S}_{+} \hat{S}_{-} 
\end{equation}  

\noindent with single-mode inhomogeneity resulting from kinetic energy of the electrons $\hbar\epsilon_{\mathbf{k}}=\hbar^2 k^2/2m$, each one with mass $m$  and $\chi_F>0$ is the effective attractive $s$-wave interaction between opposite-spin fermions~\cite{Leggett1980,Nozieres1985}.

Within mean-field theory, the pairing is described by the complex order parameter or pairing gap

\begin{equation}
\Delta
=
\chi_F
\sum_{\mathbf{k}}
\left\langle
\hat c_{-\mathbf{k},\sigma}
\hat c_{\mathbf{k},\bar{\sigma}}
\right\rangle =  \chi_F\langle \hat{S}_-\rangle
\end{equation}

\noindent that is proportional to the complex projection of the total Bloch vector onto the equatorial plane $\langle \hat{S}_-\rangle= \langle \hat{S}_x\rangle -i \langle \hat{S}_y\rangle$. The pairing interaction reduces the energy of the system by placing modes in superpositions of occupied and unoccupied, but balanced against the energy costs $\epsilon_\textbf{k}$ arising from having finite occupation of modes at higher momentum.  This competition gives rise to quasiparticles with excitation energy $\hbar B_{\mathbf{k}}
=\hbar \sqrt{(\epsilon_{\mathbf{k}}-\mu)^2+\Delta^2}$.  In the pseudospin language $2B_\mathbf{k}$ plays the role of an effective Rabi frequency. In $B_{\mathbf{k}}$ the  kinetic energy now is 
measured relative to the chemical potential $\hbar\mu$, which sets the total number of pairs.

In the BCS regime, the chemical potential remains close to the Fermi energy ($\mu>0$), and the minimum quasiparticle excitation occurs near the Fermi surface  reflecting the presence  of  extended  Cooper pairs (see Appendix~\ref{app:pairing}, and Fig.~\ref{fig:fig1_theory}a,b). In the BEC regime, the chemical potential becomes negative ($\mu<0$), indicating that the energy required to break a pair becomes dominated by the pair-binding energy.

A direct probe of the pairing character can be obtained from the non-equilibrium dynamics of the order parameter induced by a sudden quench of the interaction strength $\chi_F$.

At long times, in the so-called phase II,  the order parameter approaches a finite  steady-state value 
\begin{equation}
\Delta(t\to \infty) =
\Delta_{\infty}e^{-i2 \mu_\infty t} ,
\label{eq:deltalongtime}
\end{equation}
with $\Delta_{\infty}$ taken to be real.  Here $\mu_\infty$ is the non-equilibrium chemical potential, which sets the precession  frequency of the pairing gap  and remains close to its ground state value $\mu_\infty\sim \mu$ for weak quenches (see Ref. \cite{supplement}). It has been  also analytically   shown in linear response  \cite{Gurarie2009,supplement},
that towards its approach to steady state, $|\Delta(t)|$  features transient  oscillations with 
a power-law envelope
\begin{equation}
|\Delta(t)| =
\Delta_{\infty}
+
A t^{-\alpha}
\cos(\omega_{\mathrm{H}} t+\phi),
\label{eq:order_param}
\end{equation}
where $A$ and $\phi$ depend on the quench, $\omega_{\mathrm{H}}$ is the Higgs oscillation frequency, and $\alpha$ is the decay exponent. The oscillation frequency is set by twice the minimum quasiparticle excitation energy, $\omega_{\mathrm{H}}=2B_{\mathrm{min}}=
\min_{\mathbf{k}}\left\{
2\sqrt{(\epsilon_{\mathbf{k}}-\mu_\infty)^2+\Delta_{\infty}^2}\right\}\approx  2\chi_F  \sqrt{ |\langle \hat {S}_-\rangle| ^2+ \langle \hat {S}_z\rangle ^2 }$. 
 When the system remains in the  BCS regime,  $\mu_\infty>0$, the Higgs frequency approaches $\omega_{\mathrm{H}}\simeq2|\Delta_{\infty}|$. In contrast, in the BEC regime, when $\mu_\infty<0$,  the oscillation frequency approaches  $\omega_{\mathrm{H}}\simeq2\chi_F|\langle \hat {S}_z\rangle|$. The decay exponent $\alpha$ is likewise determined by the quasiparticle spectrum and density of states.  In   the BCS regime, the Higgs oscillations feature  a slow universal decay with exponent $\alpha=1/2$. On the other hand, in  the BEC regime the modified excitation spectrum and emergence of tightly bound pairs lead to substantially faster decay; $\alpha$ is no longer universal and strongly depends on the density of states and single particle dispersion. 

\section{Experimental System}
\begin{figure}[!ht]
\centering
\includegraphics[width=0.4\textwidth]{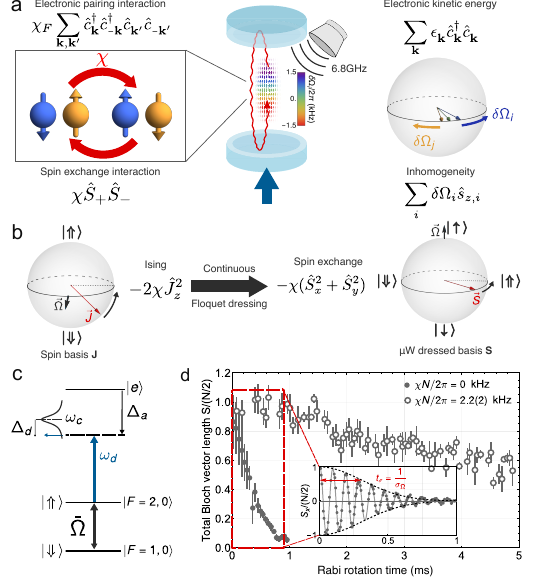}
\caption{
\textbf{Experimental overview.}\small{
\textbf{a.} The BCS Hamiltonian under Anderson pseudospin mapping is realized via cavity-mediated spin exchange interaction (left) and inhomogeneous microwave Rabi frequencies experienced by the atoms (right). \textbf{b.} The system generates all-to-all Ising interactions in spin basis $\mathbf{J}$ (left). Under continuous Floquet microwave driving, the interaction time-averages to a spin exchange interaction in the microwave-dressed ($\rm \mu W$) spin basis $\mathbf{S}$.
\textbf{c.} Relevant $^{87}$Rb atomic energy levels, dressing laser frequency $\omega_d$ (blue arrows), cavity frequency $\omega_c$ (black Lorentzian), average microwave-dressing Rabi frequency $\bar{\Omega}$, dressing laser detuning from the optical atomic transition $\Delta_a$, and average dressing laser detuning from the dressed cavity resonance $\Delta_d$. \textbf{d.} Measured collective Bloch vector length as a function of Rabi rotation time with (solid markers) and without (open markers) interactions. Without interactions, Rabi oscillations decay rapidly, with a decay envelope well-approximated by a Gaussian (inset).}
}
\label{fig:fig2_Anderson}
\end{figure}

To observe the predicted changing behavior of Higgs oscillations across the BEC-BCS crossover, we begin by laser cooling $N = 8.8 \times 10^{4}$ $^{87}$Rb atoms inside a vertically oriented, two-mirror standing-wave cavity, as shown in Fig.~\ref{fig:fig2_Anderson}a. We encode a spin-1/2 in the  two ground hyperfine states $\ket{\Uparrow}\equiv\ket{F=2,m_F=0}$ and $\ket{\Downarrow}\equiv\ket{F=1,m_F=0}$ (Fig.~\ref{fig:fig2_Anderson}c).   The non-degenerate atoms can be treated as distinguishable particles with atom label $i$. In this basis, we define single-particle spin operators $\hat{j}_{\beta,i}$ for atom $i$ and collective spin operators $
\hat{J}_\beta = \sum_{i=1}^N \hat{j}_{\beta,i} $ with  $\beta\in\left\{x,y,z,+,-\right\}$. 

To observe the power-law decay of the Higgs oscillations, it is critical that the interactions with the cavity be highly homogeneous.  To achieve this, the atoms are allowed to fall along the cavity axis, traversing from a node to an antinode in approximately 1~$\mu$s, a timescale much shorter than all other relevant dynamics. This rapid axial motion effectively time-averages each atom’s coupling to the standing-wave cavity field, thereby realizing a homogeneous atom–cavity coupling. Transitions  to other momentum states can be neglected because they are both non-resonant and not collectively enhanced.  Radial confinement to an rms transverse extent of 8.5~$\mu$m, much less than the 72~$\mu$m cavity mode waist size, is achieved by an intracavity hollow optical dipole trap as described in \cite{Thompson2022SAI}.

As a first step, we use the cavity to mediate an all-to-all Ising or one-axis twisting interaction between the atoms  $\hat{H}_{\rm Ising} / \hbar = -2\chi \hat{J}_z^2$, where the interaction depends on the population difference between  $\ket{\Uparrow}$ and $\ket{\Downarrow}$ \cite{leroux2010implementation,Vuletic2019NearUnitary,Thompson2022SAI} (see Appendix~\ref{app:hamiltonian}). The interaction strength $2\chi$ is controlled by a dressing laser applied along the cavity axis, as shown in Fig.~\ref{fig:fig2_Anderson}a.

We then use continuous Floquet driving via  microwaves to couple $\ket{\Downarrow} \leftrightarrow \ket{\Uparrow}$ causing the atoms to Rabi flop with ensemble-averaged Rabi frequency $\bar{\Omega}= 2\pi \times 10~\mathrm{kHz}$. The fast oscillations  transform the Ising interaction into the desired all-to-all spin-exchange interaction $ -2\chi \hat{J}_z^2\rightarrow -\chi\left(\hat{J}_y^2+\hat{J}_z^2\right) = -\chi\left(\hat{S}_x^2+\hat{S}_y^2\right) =- \chi (\hat{S}_+\hat{S}_- -\hat{S}_z)$. 
Our single-mode $\hat s_{_\beta,i}$ and collective $\hat{S}_\beta$  Anderson pseudo-spin operators are defined with respect to the energy eigenstates in the microwave-dressed basis $\ket{\uparrow}_i = (\ket{\Uparrow}_i + \ket{\Downarrow}_i)/\sqrt{2}$ and $\ket{\downarrow}_i = (\ket{\Uparrow}_i - \ket{\Downarrow}_i)/\sqrt{2}$.  They are explicitly related to the bare basis operators by $\hat{s}_{x,i}=-\hat{j}_{z,i}$,
$\hat{s}_{y,i}=\hat{j}_{y,i}$, and $\hat{s}_{z,i}=\hat{j}_{x,i}$.  We have also mapped the previous mode index $\mathbf{k}$ to a single-atom label $i$.

Our free-fall geometry yields nearly uniform all-to-all couplings, with only small residual inhomogeneities arising from the cloud’s finite radial temperature. 
We describe these inhomogeneities by replacing $\chi \hat{S}_+\hat{S}_- \to 
\chi \hat{S}_+\hat{S}_- + \chi \sum_{i,j} \zeta_{ij}  \hat{s}_{+,i} \hat{s}_{-,j}$ where $\chi$ is the ensemble-averaged atom-atom coupling and the inhomogeneity parameter $\zeta_{ij}$ has zero mean and a standard deviation $\ll 1$.  We thus neglect these corrections in the following  discussion,  but retain them in the numerical simulations \cite{supplement}.


 In addition to the cavity-mediated exchange interactions, we use a  linear gradient of the microwave field amplitude across the 1~mm axial extent of the atomic cloud such that atom $i$ experiences a microwave Rabi frequency,  $\Omega_i = \bar \Omega + \delta \Omega_i$. The  deviations from the average $\delta\Omega_i$ are well approximated by a  Gaussian  distribution with rms spread $\sigma_{\Omega}\approx 2\pi \times 0.8$ kHz,  confirmed by a Gaussian fit to the observed decay of Rabi oscillations (Fig.~\ref{fig:fig2_Anderson}d).

The full Hamiltonian  experienced by the atoms in a frame rotating at frequency, $\bar{\Omega} +\chi+ 2 \tilde{\mu}$, reduces to  Eq. \eqref{eq:hamiltonianBCS} if one identifies  ${\bf k}\to i$, $\chi_F\to \chi$, $\epsilon_{\mathbf{k}} \to \delta \Omega_i/2-\tilde{\mu}=\Delta \Omega_i$ (see Appendix~\ref{app:hamiltonian},
and Refs. \cite{norcia2018cavity,luo2024hamiltonian,Thompson2022SAI}).  
Here we introduced a frequency offset $2\tilde{\mu}\sim -2 \sigma_\Omega$,  which fixes the bottom of the effective kinetic energy in our simulator to zero.

\begin{figure*}[!t]
\centering
\includegraphics[width=\textwidth]{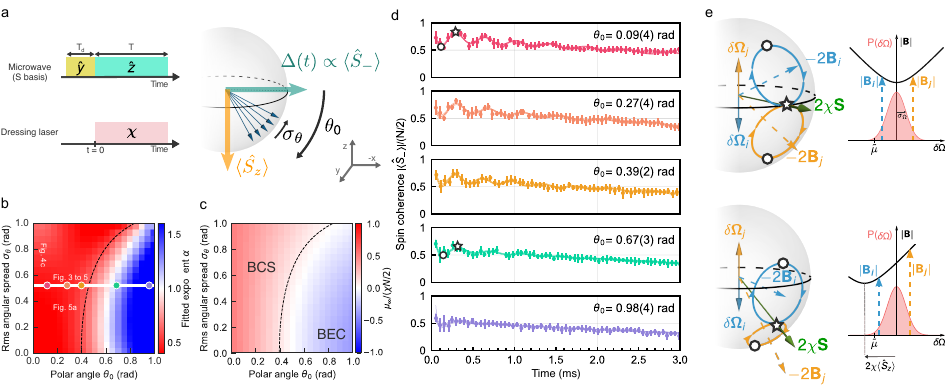}
\caption{\textbf{Quench protocol.}
\textbf{a.} We prepare initial spin textures (individual arrows $\mathbf{s}_i$) with a mean polar angle $\theta_0$ and an rms angular spread $\sigma_{\theta}$, which together determine the total magnetization $\langle \hat{S}_z \rangle$. The collective transverse component $\langle \hat{S}_- \rangle$ maps to the complex pairing gap $\Delta(t) \propto \langle \hat{S}_-(t) \rangle$. The experimental sequence for state preparation and the quench at $t=0$ are shown in the left panel. 
\textbf{b.} Numerically calculated phase diagram of the Higgs oscillation damping exponent $\alpha$ as a function of $\theta_0$ and $\sigma_\theta$, for the  experimentally relevant  parameter regime. 
The lines and labels show experimentally explored cuts through the phase diagram.
\textbf{c.} Numerically calculated  non-equilibrium chemical potential, $\mu_{\infty}$, as a function of $\theta_0$ and $\sigma_\theta$ (Ref.~\cite{supplement}).   The sign of $\mu_{\infty}$ can be associated with the  BCS and BEC regimes. The dashed boundary indicates $\mu_{\infty}=0$.  The correlation between $\alpha=0.5$ and $\mu_{\infty}>0$  and between $\alpha>0.5$ and $\mu_{\infty}<0$ is clearly visible in panels b and c. 
\textbf{d.} Time dynamics of measured $|\langle \hat S_-(t) \rangle|/(N/2) \propto |\Delta(t)|$ after the quench at $t=0$ for several initial tipping angles $\theta_0$ of the spin texture. \textbf{e.} In both the BCS (top) and BEC (bottom) regimes, the transverse component of the collective Bloch vector $\mathbf{S}$ oscillates in time as individual Bloch vectors $\mathbf{s}_i$ (blue orbit) and $\mathbf{s}_j$ (yellow orbit) precess around distinct effective magnetic fields $\mathbf{B}_i$ (blue dashed arrow) and $\vec{B}_j$ (yellow dashed arrow).  Damping of the Higgs oscillations arises from inhomogeneity in their precession frequencies $2|\mathbf{B}_i|$ causing relative dephasing of their orbits (solid color circles).  When $\mathbf{S}$ is tilted away from the equator (bottom), the inhomogeneity in $2|\mathbf{B}_i|$ across the ensemble is larger across the mode density distribution ($P(\delta\Omega)$ in pink) than when $\mathbf{S}$ lies on the equator (top), leading to faster damping of the oscillations.}
\label{fig:fig3}
\end{figure*}

The spin-exchange interaction can be written as $\hat{S}_+ \hat{S}_- = \hat{\mathbf{S}}\cdot\hat{\mathbf{S}} - \hat{S}_z^2+ \hat{S}_z$. For convenience  we further transform to a  rotating frame where  the collective Bloch vector $\langle  {\bf \hat S}\rangle$ is stationary. This frame rotates   about $\hat{z}$ at a frequency  approximately equal to $  2\chi \langle\hat S_z\rangle+ \bar \Omega+\chi$ (see Ref. \cite{supplement}).
In this rotating frame, at the mean-field level which neglects quantum correlations, we have removed the Ising interaction $\chi \hat S_z^2$, and only retained the Heisenberg term $-\chi \hat{\mathbf{S}}\cdot\hat{\mathbf{S}}$. The resulting mean-field dynamics is a simple   precession  of each spin $i$ around  a fictitious  magnetic field $ {\bf B}_i$ arising from the combined action of a collective self-generated  field     plus  the   single-atom inhomogeneous   microwave drive Rabi frequency (see Fig. \ref{fig:fig3}e):

\begin{equation}
\hat{H}_{\rm MF}/\hbar = -2 \sum_i {\bf B}_i\cdot {\bf \hat s}_i,\label{eq:MFBCrot1}\quad 2{\bf B}_i = 2 \chi \langle \hat{\bf {S}}\rangle -\delta \Omega_i \hat{z}
\end{equation}

We first verify the emergence of the exchange Hamiltonian through the enhancement of coherence time as studied in related systems~\cite{norcia2018cavity,Schleier-Smith2020GapProtection,ThompsonMomentumExchange2024,Thompson2025OpticalGap}. As shown in Fig.~\ref{fig:fig2_Anderson}d and its inset, the Rabi oscillations of the different atoms occur at different frequencies $\delta \Omega_i$ leading to rapid decay of the collective Bloch vector length within $\lesssim 1~\mathrm{ms}$.  In comparison, when the dressing laser is applied to generate  spin-exchange interactions, the observed Rabi oscillations persist for much longer (open points) since the collective term $\chi\hat{\bf S}\cdot \hat{\bf S}$ opens an energy gap between states of different total $S$ that pushes the coupling to states of smaller $S$ off-resonance, thereby preserving the Bloch vector length~\cite{norcia2018cavity,Schleier-Smith2020GapProtection}.

\section{Experimentally observed Higgs oscillations}

\subsection{Spin-texture preparation}

To probe the dynamical behavior of the Higgs mode across the BEC--BCS crossover, we  prepare initial pseudospin textures with a controllable overlap with the mean-field ground state of Eq.~\eqref{eq:MFBCrot1} (Fig.~\ref{fig:fig3}a and Appendix~\ref{app:spin_texture}). The spin textures will be characterized by a mean tipping angle $\theta_0$, measured from the equator of the dressed-state Bloch sphere, and an rms angular spread $\sigma_\theta$ of the spin tipping angles  about the mean (Fig.~\ref{fig:fig4}a).  Varying the mean tipping angle \(\theta_0\) allows us to tune the prepared state to lie either in the BCS or BEC regime of Eq.~\eqref{eq:MFBCrot1},
in a way conceptually similar to controlling the interaction strength through a Feshbach resonance in ultracold atomic gases~\cite{cornish2000stable}.  Specifically, we will identify $\theta_0=0$ with the BCS regime, crossing over to the BEC regime as $\theta_0$ tips below the equator.  The crossover angle is when the chemical potential $\mu\sim \mu_\infty$ becomes negative, which in the spin model here, corresponds to the angle at which $|\chi  \langle \hat S_z\rangle|\sim |\tilde \mu|$. The parameters \(\theta_0\) and \(\sigma_\theta\) can  be tuned largely independently (See Appendix~\ref{app:spin_texture}, and Ref~\cite{supplement}). 

\subsection{Higgs oscillations across BEC-BCS }

After the spin-texture is initialized, the laser generating the exchange interaction with strength \(\chi\) is turned on (Fig.~\ref{fig:fig3}a), inducing damped Higgs oscillations.  




 

\begin{figure*}[!t]
    \centering
    \includegraphics[width=\textwidth]{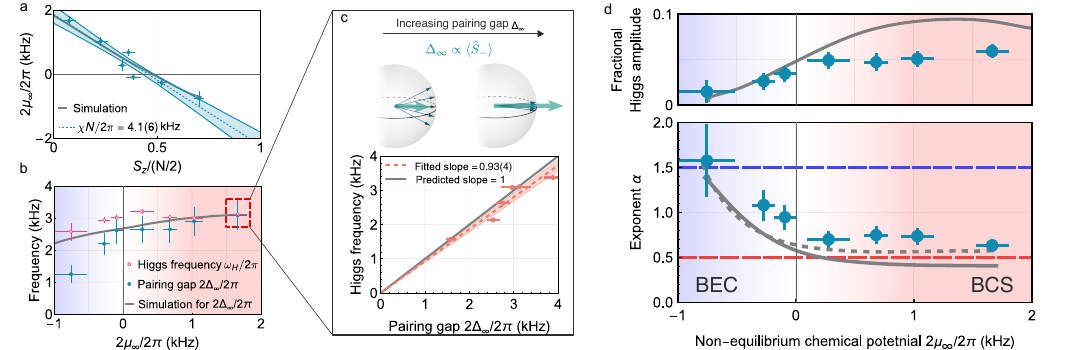}
    \caption{
    \textbf{Higgs oscillations across the BCS--BEC crossover.}
    \textbf{a.} Measured non-equilibrium chemical potential $\mu_\infty$ as a function of initial population difference $\langle \hat{S}_z \rangle/(N/2)$. 
    \textbf{b.} Measured pairing gap $2\Delta_\infty$ (blue solid markers) and Higgs oscillation frequency (red open markers) as a function of the non-equilibrium chemical potential $2\mu_\infty$.
    \textbf{c.} Measured Higgs frequency $\omega_H/2\pi$ in the BCS regime as a function of the pairing gap. The pairing gap was varied by adjusting the initial angular spread $\sigma_\theta$, which effectively controls the length of the collective Bloch vector $|\langle \mathbf{\hat{S}}\rangle |$. The fitted slope (red with 1 s.d. uncertainty band) is in close agreement with the expected Higgs frequency  scaling $\omega_\mathrm{Higgs} = 2\Delta_\infty$ (grey solid line).
    \textbf{d.} Fitted Higgs amplitude  (top) and  decay exponent $\alpha$ (bottom) as a function of the non-equilibrium chemical potential $2\mu_\infty$, showing the crossover from the BCS to the BEC regime. The grey dashed curve in the lower panel shows the numerical simulation
including a finite radial temperature corresponding to an rms transverse
extent of $\sigma_x=\sigma_y=8.5(5)~\mu\mathrm{m}$~\cite{supplement}. For all panels, grey solid curves represent theoretical calculations based on the full atom-cavity model, including the relevant sources of decoherence but not the
effect of finite radial temperature \cite{supplement}.}
    \label{fig:fig4}
\end{figure*}

The  directly observed  Higgs oscillations are shown  in Fig.~\ref{fig:fig3}{d}, where we show time  traces of the collective spin coherence $|\langle \hat S_-(t)\rangle|$,  which is directly proportional to the pairing gap \begin{equation}|\Delta(t)|=\chi |\langle\hat{S}_-(t) \rangle|= \chi \sqrt{\langle \hat{S}_x(t)\rangle^2+ \langle \hat{S}_y(t)\rangle^2},\end{equation} for several initial tipping angles $\theta_0$ with trace color corresponding to the corresponding points in the dynamical phase diagram of Fig.~\ref{fig:fig3}{b}.

The phase diagrams  (Fig.~\ref{fig:fig3}b,c)  are  obtained  by  performing  numerical simulations where we plot the decay exponent  $\alpha$,  and $\mu_\infty$ extracted  from the ideal dynamics (pure Hamiltonian evolution) over the accessible parameter regime of our simulator. Experimental exploration of the dynamical phase diagram along  horizontal cuts at fixed $\sigma_\theta$  is shown in both Fig. \ref{fig:fig4}d and \ref{fig:fig5}a.

The Higgs data show transient oscillations that damp as a function of time.    
As the initial tipping angle $\theta_0$ is increased, we observe  the Higgs oscillations damp faster and fewer oscillations can be seen.  This trend  is the key qualitative result: the damping of the Higgs oscillations increases as one moves from the BCS to BEC limit.

We also show fits (solid lines) in Fig~\ref{fig:fig3}{d} that incorporate the power-law decay of the Higgs oscillations as well as a much slower exponential decay of the order parameter largely arising from free-space scattering (see Appendix~\ref{app:fit} and  Ref. \cite{supplement}). Each plotted data point in Fig.~\ref{fig:fig3}{d} represents an average over half of a Rabi cycle (approximately 10 points) for visualization, while all fits are performed on the individual data points.

\subsection{Non-equilibrium chemical potential and pairing energy gap}

From the  fits to the data  we extract the Higgs oscillation frequency $\omega_H$ and 
the long-time value of the measured pairing gap $|\Delta_{\infty}|$ by removing the slow exponential decay due to free-space scattering.  We can also determine complex pairing gap precession  frequency associated to the  non-equilibrium chemical potential  (Eq.~\eqref{eq:deltalongtime}):
\begin{equation}\mu_\infty \approx  \chi \langle S_z\rangle-\tilde{\mu}\end{equation}  We measure this frequency via the observed change in the frequency of the microwave-induced Rabi flopping (in the original laboratory frame) when the interaction (i.e. the dressing laser) is applied. The difference in oscillation period $\chi \Delta T=\pi/ \langle S_z\rangle$,  with and without interactions allows us to extract $\mu_{\infty}$.


In Fig.~\ref{fig:fig4}a, we present the measured $2\mu_{\infty}$ versus total spin projection $\langle \hat S_z\rangle$ which is varied by changing the spin texture's tipping angle $\theta_0$. The grey solid curve shows numerical simulations of the full atom–cavity model including dissipation. The observed variation largely reflects the mean-field one-axis twisting shift $2\chi \langle \hat{S}_z\rangle$ arising from the exchange interaction with the subleading correction in Ref.~\cite{supplement}. 

In Fig.~\ref{fig:fig4}b, we combine measurements when varying $\theta_0$ to show how  $\Delta_\infty$ (blue points)  and  $\omega_H$ (red open points) vary  with  $\mu_\infty$. The measured observables agree with  the numerical simulation (grey solid curve) except for an outlier at $2\mu_\infty/2\pi = -0.75$ kHz at large tipping angle.  

In Fig.~\ref{fig:fig4}{c}, we compare  $\omega_H$ to  $\Delta_\infty$ in the BCS regime ($\theta_0\approx 0$). Here, $\Delta_\infty$ is tuned by varying the initial rms angular spread $\sigma_\theta$, which effectively controls the transverse component of the Bloch vector. The data is well described by  $\omega_H = 1.86(7) \Delta_\infty$, in good agreement with the expected scaling for Higgs oscillations $\omega_H = 2 \Delta_\infty$.




\subsection{Power-law decay}

To quantitatively characterize the Higgs oscillation dynamics across the different regimes,
we extract the Higgs decay exponent $\alpha$ and oscillation amplitude by fitting the measured
collective Bloch-vector length $S(t)=|\langle\hat{\mathbf S}(t)\rangle|$.
For presentation, the data in Fig.~\ref{fig:fig3}d are shown as the transverse spin coherence
$|\langle\hat S_-(t)\rangle|$ (see Appendix~\ref{app:fit}). The results are plotted versus the non-equilibrium chemical potential $\mu_{\infty}$ obtained by fitting the experimental data in Fig.~\ref{fig:fig4}{a}  at the corresponding tipping angle $\theta_0$.  We observe a robust plateau of $\alpha \approx 0.65(8)$ for initial states prepared near the equator (Fig.~\ref{fig:fig4}d, bottom).  This value is  close to but  higher than the ideal $\alpha=0.5$ value. We attribute this discrepancy to the small interaction inhomogeneities $\zeta_{ij}$ arising from the finite radial temperature of the atoms. Accounting for the radial temperature in numerical simulations yields the prediction $\alpha=0.58$ (see Ref.~\cite{supplement}), consistent with the experimentally measured value. The uncertainty on the reported $\alpha$ reflects the spread in measured values and variation depending on the data analysis, such as the included fit range.
As the non-equilibrium chemical potential decreases (i.e. the tipping angle $\theta_0$ increases) we cross toward the BEC regime and one sees that the power-law decay exponent $\alpha$ increases smoothly.
The crossover at  $\mu_{\infty}\approx 0$  is indicated by the grey vertical line in Fig.~\ref{fig:fig4}d. 

The experimental data are in good agreement with the numerical simulation (grey lines) of the full atom-cavity model that accounts for known relevant sources of decoherence  including free-space scattering and cavity loss  \cite{supplement}.
The slightly faster damping observed in the BCS regime is captured by simulations that account for finite radial temperature (Fig.~\ref{fig:fig4}d dashed curve) \cite{supplement}.


Fig.~\ref{fig:fig3}{e} provides qualitative insight into \emph{why} the damping of the Higgs oscillations changes at all as one moves from the BCS to BEC regimes.  Following a quench, the pairing gap  emerges from the coherent sum of pseudospins that  start orbiting  about distinct axes (example orange and blue orbits for spins $i$ and $j$ about $2\mathbf{B}_{i}$ and $2\mathbf{B}_{j}$) with a precession frequency  set by  $2|\mathbf{B}_i|=2\sqrt{ \Delta_\infty^2+(\mu_\infty-\Delta \Omega_i)^2}=2\sqrt{ \Delta_\infty^2+(\chi \langle \hat S_z\rangle-\delta \Omega_i/2)^2}$.  
At long times, small differences in the effective Rabi frequencies accumulate into large differences in the phase of their orbits.  This dephasing of the orbits produces the damping of the Higgs oscillations.

The damping rate in the two regimes is governed by the location of the stationary (saddle) point $(\delta \Omega_{i_{\rm sad}}/2-\chi \langle \hat S_z\rangle )=0$ in the spectrum of effective Rabi frequencies.  In the BCS regime ($\theta_0\sim0$ or $\langle \hat S_z\rangle\sim 0$) the saddle point lies within the density of states  $ |\delta \Omega_{i}|/\Delta_\infty \ll 1$,  and  therefore   the excitation spectrum varies as $|\mathbf{B}_i|\approx \Delta_\infty + (\delta \Omega_{i})^2/(8|\Delta_\infty|)$ leading to a slow dephasing of the orbits. In the deep  BEC regime where $ \langle \hat S_z\rangle  \sim - N/2$, the stationary point no longer lies within the bulk of the mode density but instead approaches towards the tails.  In this case, it is better to expand  around the peak of the mode density where the  effective Rabi frequency varies linearly as   $|\mathbf{B}_i|\approx |\chi \langle \hat S_z\rangle-\delta \Omega_{i}/2| $ resulting in a broader distribution of precession frequencies and much more rapid dephasing. 

A detailed analytical discussion is provided in Ref.~\cite{supplement}, where approximate solutions are derived in the deep BCS and deep BEC limits. While the BCS regime exhibits the robust asymptotic power-law decay $t^{-1/2}$, the Gaussian mode distribution considered here leads to $|\alpha|>1$. In practice, however, the long-time oscillations in the BEC regime are strongly suppressed and become difficult to resolve both experimentally and numerically. Consequently, throughout this work the effective decay exponent on the BEC side is extracted by fitting the first few oscillation peaks, which adequately capture the experimentally accessible dephasing dynamics and provide a consistent comparison when tuning from the BCS to BEC regime  (Fig.~\ref{fig:fig4}d).

\subsection{Universality and robustness}
\begin{figure}[!t]
\centering
\includegraphics[width=0.5\textwidth]{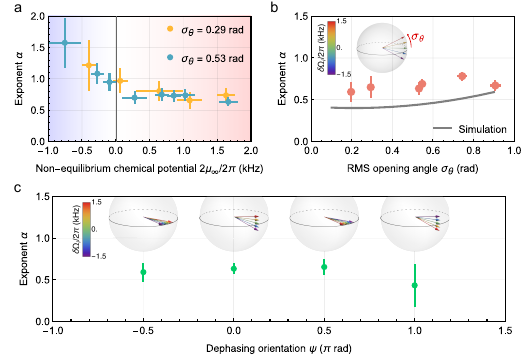}
\caption{
\textbf{Universality and robustness of Higgs dynamics.}
\textbf{a.} Fitted decay exponent $\alpha$ of the Higgs oscillations as a function of measured non-equilibrium chemical potential $2\mu_\infty$. Data are shown for two spin textures with rms angular spreads $\sigma_\theta=0.29$ and $0.53~\mathrm{rad}$, confirming that the power-law decay behavior is largely insensitive to the initial spreading of the spin texture.
\textbf{b.} Decay exponent $\alpha$ versus $\sigma_\theta$ in the BCS regime $\theta_0=0$. The measured exponent (red points) remains approximately constant with a slight increase for larger $\sigma_\theta$, in qualitative agreement with numerical simulations (solid grey curve).
\textbf{c.} Decay exponent $\alpha$ versus the orientation of the spin texture's spreading in the BCS limit $\theta_0=0$.  The point at $\psi=0$ corresponds to the orientation of the spin texture associated with a BCS equilibrium ground state. The Bloch-sphere insets illustrate the corresponding orientations of the spin-texture plane. The approximately constant value of $\alpha$ indicates that the Higgs dynamics are insensitive to the orientation of the initial dephasing. 
}
\label{fig:fig5}
\end{figure}

A central prediction of the quench dynamics of  an $s$-wave BCS–BEC superfluid  is  the universality of $\alpha=1/2$ in the BCS regime. We  test this universality by varying both the angular spread $\sigma_\theta$ of the spin texture and its orientation $\psi$ of its plane relative to the mean Bloch vector.


In Fig.~\ref{fig:fig5}a, we plot the fitted decay exponent $\alpha$ as a function of the measured $\mu_{\infty}$ for two different spin textures with spread $\sigma_\theta = 0.29$ rad (orange points) and $\sigma_\theta = 0.53$ rad (blue points), corresponding to the two horizontal cuts in the phase diagram in Fig.~\ref{fig:fig3}b.  Conceptually, the two spin textures with smaller and larger spreads correspond to having prepared the ground state of two different BCS Hamiltonians in which the interaction strength $\chi$ is larger and smaller, respectively. 

The power-law decay exponents of both spin textures exhibit the same trend as the system is tuned from the BCS to the BEC regime. Comparing $\alpha$ at matching values of the non-equilibrium chemical potential $\mu_\infty$ between the two textures gives an average ratio of $\alpha_{\sigma_\theta=0.53}/\alpha_{\sigma_\theta=0.29}=0.93(7)$ across  the BCS regime $\mu_\infty\ge 0$.


In Fig.~\ref{fig:fig5}b, we explore the insensitivity to the initial spin texture spreading by showing the measured $\alpha$ versus spreading $\sigma_\theta$ in the BCS limit, i.e.~at  $\theta_0=0$.  This corresponds to exploring the vertical cut in the phase diagram of Fig.~\ref{fig:fig3}b.  The numerical simulation (grey curve)  predicts that the decay exponent $\alpha$ should remain relatively constant versus $\sigma_\theta$.   Consistent with this prediction, the measured $\alpha$ remains  approximately constant over the explored range.


Finally, in Fig.~\ref{fig:fig5}{c} we go beyond the accessible spin textures that can be prepared by relaxing to the equilibrium ground state of a BCS Hamiltonian.
 BCS equilibrium ground states correlate the tipping angle of a mode's spin vector with its single-particle energy.  We define this orientation of the plane of the spin-texture as $\psi=0$.  We use coherent microwave control during the spin-texture preparation to also prepare initial spin textures $\psi=-\pi/2, 0, \pi/2, \pi$, as illustrated by the inset Bloch spheres.  Again, the fitted exponent $\alpha$ remains approximately unchanged across all orientations, indicating that the Higgs decay dynamics are insensitive to the microscopic detail of the orientation of the spin-texture's spread.


\section{Conclusions}
In this work, we have realized quench dynamics in a controllable cavity QED platform that emulates key features of fermionic superfluids across the BEC to BCS crossover. By tracking the time evolution of the order parameter $\Delta(t)\propto\langle \hat{S}_-(t)\rangle$, we identify a regime in which pairing persists and exhibits damped Higgs oscillations. The observed oscillation frequencies and damping dynamics evolve across the crossover in a manner consistent with long-standing theoretical predictions for non-equilibrium pairing dynamics \cite{Yuzbashyan2006,Yuzbashyan2006PRB,Gurarie2009} that had not been observed previously. In particular, we observe slower power-law relaxation in the BCS regime and faster decay in the BEC regime, reflecting the changing quasiparticle spectrum $|\mathbf{B}_i|$ underlying the collective dynamics. While quantitative deviations from idealized theoretical models remain (measured $\alpha=0.65(8)$ compared to ideal theory $\alpha=0.5$, and numerical value at finite temperature  $\alpha=0.58$ \cite{supplement}), our results provide the first experimental observation of these predicted dynamical behaviors and establish cavity QED systems as a powerful platform for studying far-from-equilibrium many-body pairing phenomena.

An important next step is to extend these studies to systems with nontrivial topology, such as $p+ip$ and $d+id$ superconductors \cite{Chu2025arXiv}. In contrast to the smooth crossover explored here, these systems are expected to exhibit genuine dynamical phase transitions associated with changes in topological order. Theoretical work has predicted rich non-equilibrium phase structure and topological transitions in quenched chiral superconductors \cite{Foster2013,Foster2013PRB,Dzero2007PRL}, while recent proposals show that such pairing symmetries can be engineered and probed in cavity QED platforms. Implementing controlled quenches in these systems offers a promising route to directly observe dynamical topological transitions that have so far remained experimentally elusive.


Furthermore, our platform opens the door to probing transport-related phenomena in superfluids in a controlled manner. Effects such as Andreev reflection and coherent transport across mesoscopic barriers, which have been observed and quantitatively modeled in ultracold atom experiments \cite{Husmann2015Science,Lebrat2018PRX,huang2023superfluid}, could be investigated here with direct control over interactions and non-equilibrium protocols. 

\begin{acknowledgments}

This material is based upon work supported by the US Department of Energy, Office of Science, National Quantum Information Science Research Centers, Quantum Systems Accelerator, the Vannevar-Bush Faculty Fellowship, and the Simons Collaboration on Ultra-Quantum Matter, Simons Foundation grant (651440). We acknowledge additional funding support from the NSF under grant nos. 1734006 (Physics Frontier Center) and NSF QLCI Award -2552714 (Q-SEnSE); and NIST. We acknowledge helpful feedback on the manuscript from Tim de Jongh and Leo Radzihovsky.
\end{acknowledgments}

\bibliography{BCS}

\appendix
\section{ Pairing  across the BEC-BCS crossover}
\label{app:pairing}
Within the one-channel description, the BEC--BCS crossover is encoded in the structure of the paired state, its excitation spectrum--- 
in the frame where the single-particle energy  is measured relative to the chemical potential $\mu$---, and the momentum occupation distribution described by   $
N_{\mathbf{k}}
=\frac12\left(
1+\frac{\mu-\epsilon_{\mathbf{k}}}{B_{\mathbf{k}}}
\right)$, or equivalently, $\langle \hat{s}_\textbf{k}^z\rangle  = (\mu-\epsilon_\textbf{k})/(2B_\textbf{k})$.

In the BCS regime, the chemical potential remains close to the Fermi energy ($\mu>0$), and the minimum quasiparticle excitation occurs near the Fermi surface (Fig.~\ref{fig:fig1_theory}a,b), reflecting the presence  of  extended  Cooper pairs. Correspondingly, the momentum occupation retains a clear remnant of the Fermi surface: low-energy momentum states are predominantly occupied, while high-energy states remain mostly empty, with pairing only smoothing the distribution near the Fermi energy. The condition $N_{\mathbf{k}}=1/2$, or equivalently $\langle \hat{s}_\mathbf{k}^z\rangle =0$ defines a remnant Fermi surface.  As shown in Fig.~\ref{fig:fig1_theory}d, the pseudo-spin forms a texture in which the spin orientation rotates continuously from up to down, passing through the equator near the non-interacting Fermi surface.

As the interaction strength $\chi_F$ increases, the pair wavefunction broadens in momentum space and becomes increasingly localized in real space, signaling the formation of tightly bound composite pairs. In the BEC regime, the chemical potential becomes negative ($\mu<0$), indicating that the energy required to break a pair exceeds the energy of free fermions (see Fig.~\ref{fig:fig1_theory}a,b). 
Correspondingly, the minimum quasiparticle excitation shifts from the vicinity of the Fermi surface to the bottom of the single-particle dispersion, $\epsilon_{\mathbf{k}}=0$,  where $B_{\min}=\sqrt{\mu^2+\Delta^2}$ and becomes dominated by the pair-binding energy. 
At the same time, the momentum occupation broadens substantially and the remnant Fermi surface disappears, such that no momentum state satisfies $N_{\mathbf{k}}=1/2$.  In the pseudo-spin picture, the spin-texture starts below the equator and smoothly rotates towards down such that $\langle \hat{s}_\mathbf{k}^z\rangle <0$ for all modes. The crossover therefore corresponds to a continuous evolution from extended Cooper pairs to tightly bound  pairs within a purely fermionic description.
\section{Experimental Realization of BCS Hamiltonian}
\label{app:hamiltonian}

To realize the BCS Hamiltonian, we use the cavity to mediate an all-to-all Ising or one-axis twisting interaction between the atoms  $\hat{H}_{\rm Ising} / \hbar = -2\chi \hat{J}_z^2$, where the interaction depends on the population difference between the two hyperfine levels $\ket{\Uparrow}$ and $\ket{\Downarrow}$ \cite{leroux2010implementation,Vuletic2019NearUnitary,Thompson2022SAI}.  The interactions are realized by driving the optical cavity   at frequency $\omega_d$ (see Fig.~\ref{fig:fig2_Anderson}a-c) establishing an intracavity coherent state with average photon number $|\beta|^2 \approx 2\times 10^3$ (Fig.~\ref{fig:fig2_Anderson}d) and $4\times10^3$ (Fig.~\ref{fig:fig3}-\ref{fig:fig4}). The  laser driving the cavity  is detuned by $\Delta_{a}= \omega_d-\omega_a = -2\pi \times 6.289$ GHz  from the $\ket{\Uparrow}$ to optical excited state $\ket{e}\equiv \ket{5^2 P_{3/2}, F'=3}$ transition $\omega_a$ at 780~nm, as illustrated in Fig.~\ref{fig:fig2_Anderson}c. The detuning $\Delta_a$ is large compared to all other relevant frequency scales including the excited state decay rate $\Gamma=2 \pi \times 6$ MHz and the cavity power decay rate $\kappa = 2\pi \times 56(3)$ kHz.  The dressed cavity mode at frequency $\omega_c$ (i.e. its frequency after accounting for an average shift due to the atoms in the cavity) is detuned from the laser driving the cavity by $\Delta_d = \omega_d-\omega_c= -2\pi \times 1.9$~MHz. This hierarchy of frequency scales permits adiabatic elimination of the optical hyperfine states $F'=3,2,1$ followed by adiabatic elimination of the cavity field to arrive at the Ising interaction between the ground states. Conceptually, the cavity resonance frequency is shifted by the spin-projection $\langle \hat J_z\rangle $, which then changes the circulating  power in the cavity, which then creates a differential AC Stark or light shift.  The parameter $2 \chi$ sets the interaction strength and is given by $2\chi = -g_{\rm eff}^2 / \Delta_d > 0$. Here $g_{\rm eff} \equiv \frac{g^{2}|\beta|}{3}\left(\frac{1}{\Delta_{a}} - \frac{1}{\Delta_{a} - \omega_{\rm hf}}\right)$, where $2g = 2\pi \times 0.96~\mathrm{MHz}$ is the single-atom vacuum Rabi splitting at a cavity antinode for the $D_2$ transition.

We then use continuous Floquet driving to transform the all-to-all Ising interaction into the desired all-to-all spin-exchange interaction. A microwave drive at $\omega_m=\omega_{\rm hf}$ is applied to resonantly couple the hyperfine levels $\ket{\Downarrow} \leftrightarrow \ket{\Uparrow}$.  The microwave transition frequency $\omega_{\rm hf}=2\pi\times 6.834$~GHz is first-order insensitive to magnetic fields.  When the dressing laser is applied it induces an average AC Stark shift of the clock transition.  We compensate for this average frequency shift  by jumping the microwave frequency to maintain microwave resonance throughout the interaction periods.  The microwave coupling causes the atoms to Rabi flop with ensemble-averaged Rabi frequency $\bar{\Omega}= 2\pi \times 10~\mathrm{kHz}$.  This transverse applied field $\bar \Omega \hat{J}_x$ causes the collective Bloch vector $\hat{\mathbf J}$ to rapidly precess about $\hat{x}$ (See Fig. \ref{fig:fig2_Anderson}b).  When the interaction is averaged over the transverse directions, it generates  a time-averaged  interaction $2\chi\hat{J}_z^2 \to \chi(\hat{J}_z^2 + \hat{J}_y^2)$.


\section{Spin texture initialization} 
\label{app:spin_texture}
The spin-texture preparation is accomplished by initializing the atoms in $\ket{\Uparrow}$ and applying the microwave rotation about the \(\hat y\) axis in the bare-spin basis ($J$-basis) for some time $T_d$, followed by a sudden \(\pi/2\) phase jump of the microwave drive (Fig.~\ref{fig:fig3}a).  This final microwave drive phase defines the microwave-dressed basis ($S$-basis) for the rest of the experiment. In the $S$-basis, the tipping angle is $\theta_0 = \bar \Omega T_d \mod{2 \pi } $ and the spread in spins is  
$\sigma_\theta=\sigma_\Omega T_d$ with the Bloch vectors all lying in the \(x\)--\(z\) plane, thus realizing a controllable overlap with the mean-field ground state of Eq.~\eqref{eq:MFBCrot1}.

\section{Fit function}
\label{app:fit}
In the absence of dissipation, the pairing gap $|\Delta(t)|=\chi|\langle\hat{S}_-(t)\rangle|= \chi S_-(t)$ is expected to oscillate with an amplitude that follows the power-law decay described in Eq.~\eqref{eq:order_param}.  However, free-space scattering of dressing laser photons causes the total Bloch vector length $S(t)=|\langle \hat{\mathbf{S}}(t)\rangle|$ and spin coherence $S_-(t)$ to decay exponentially, as seen in the measured $S_-(t)$ in Fig.~\ref{fig:fig3}d with fitted decay constant $\tau= 5.5(7)$~ms. As a result $S_-(t)$ varies at two distinct time scales:  an overall slow decay and a much faster Higgs oscillation about this slowly varying background. We therefore separate the dynamics into a slowly varying component and a rapidly oscillating component, which we write as $\frac{S_-(t)}{N/2}=A_-(t)\left(1+\frac{\eta(t)\cos\left({\int^t_0\omega_H(t') d t'}\right)}{t^\alpha}\right)$. Here, $A_-(t)$ denotes the slowly varying background value of the normalized transverse coherence about which the Higgs oscillations occur. 
In the absence of dissipation, $A_-(t)$ is constant and the integrated phase  of the Higgs oscillation is just $\omega_H t$. In the experiment, however, free-space scattering  causes the spin coherence amplitude to decay as $A_-(t)=A_-(0)e^{-t/\tau}$.


 This same decay also causes the effective interaction strength $\chi S(t)$ to decay with time~\cite{supplement}. This has an effect on the dynamics of $\omega_H(t)$ and $\eta(t)$. First, the instantaneous Higgs frequency chirps as $\omega_H(t) \approx\chi S(t) \approx e^{-t/\tau}\omega_H(0)$. Second, the fractional amplitude of the Higgs oscillation $\eta(t)$ scales as $\eta(t)\propto\frac{\sigma_\Omega}{\chi S(t)}$. As a result, in the presence of free-space scattering, $\eta(t)$ scales as $\eta(t) \propto \frac{\sigma_\Omega}{\chi S(0)e^{-t/\tau}}$. The exponential decay of $A_-(t)$ and the exponential growth of $\eta(t)$ cancel, leaving the absolute amplitude of the Higgs oscillation $A_-(t)\eta(t)/t^\alpha=A_-(0)\eta(0)/t^\alpha$ proportional to $1/t^{\alpha}$.

Combining these effects, we model the order parameter dynamics as a damped, chirped oscillation
$
\frac{S_-(t)}{N/2} = A_-(0) \left(e^{-t/\tau} + \eta(0)\frac{\cos\left[e^{-t/\tau}\omega_H(0) \tau + \phi_H\right]}{t^{\alpha}}\right)
$, where the phase $\phi_H$ is set by the details of the quench.
 For technical reasons~\cite{supplement}, it is easier to monitor the normalized collective Bloch vector length $S(t)$
 rather than $S_-(t)$.
 Since $\langle \hat{S}_z(t)\rangle$ is set by the known initial tipping angle $\theta_0$ and is conserved up to free-space scattering~\cite{supplement}, we fit the measured $S(t)$ using 
$
\frac{S(t)}{N/2} = A \sqrt{ \left(e^{-t/\tau}\sin{\theta_0}\right)^2+\cos^2{\theta_0}\left(e^{-t/\tau} +  \frac{\eta(0)\cos \phi(t)}{t^{\alpha}}\right)^2 }.
$
Here, $A$ is the normalized total Bloch-vector amplitude, and $\phi(t)$ is given as, $\phi(t) = e^{-t/\tau}\omega_H(0) \tau + \phi_H$. In this expression, using the same $\tau$ for the free-space-scattering decay
of both the transverse and longitudinal components is a good approximation
because their calculated decay rates differ by only approximately $10\%$~\cite{supplement}. Since $\theta_0$ is controlled and $A$ and $\tau$ are determined from the fit, we estimate $\langle \hat{S}_z(t) \rangle = A e^{-t/\tau}\sin{\theta_0}$ and determine the pairing gap using  $|\Delta(t)| = \chi \sqrt{S(t)^2-\langle \hat{S}_z(t) \rangle^2}$.

\begin{figure}[!t]
    \centering
    \includegraphics[width=\linewidth]{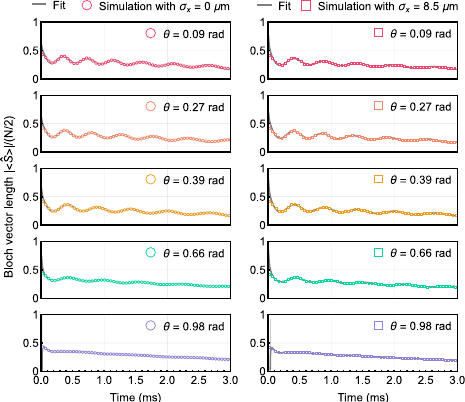}
    \caption{\textbf{ Comparison of the fit model and numerical simulations at both zero and finite radial temperatures.} Each panel shows the total Bloch vector length $|\langle\hat{S}\rangle|/(N/2)$ versus time. Left panels: numerical simulations of the full atom-cavity model (circles) fitted to the damped, chirped oscillation model (solid lines) for initial tipping angles $\theta_0 = 0.09, 0.27, 0.39, 0.66,$ and $0.98$ rad. Right panels: 
    numerical simulations and fits that include the  inhomogeneous spin–spin interactions $\zeta_{ij}$ arising from the cloud’s finite radial temperature. The  cloud’s radial temperature and an rms transverse extent $\sigma_x=\sigma_y =8.5~\mu\mathrm{m}$ were calibrated from the measured dephasing of Ramsey contrast (see \cite{supplement}).   \label{fig:num_exp_comparison}
      }
 \end{figure}

To assess the accuracy of the fitting model, we use it to fit numerical simulations of the full atom–cavity dynamics, see Fig.~\ref{fig:num_exp_comparison}, left panels. The excellent agreement shows that the fitting model accurately captures the simulated dynamics across the parameter range considered. In the right panels of Fig.~\ref{fig:num_exp_comparison}, we also show that the model accurately describes the numerical simulations that include inhomogeneous spin–spin interactions arising from the cloud’s finite radial temperature \cite{supplement}. The exponents extracted from traces such as those in Fig.~\ref{fig:num_exp_comparison} are used to calculate the solid and dashed curves shown in Fig.~4 of the main text.


\clearpage
\onecolumngrid
\clearpage
\begin{center}
{\Large\bfseries Supplementary Material}\par\vspace{0.7em}
{\large Experimental observation of Higgs oscillations and their power-law decay across pairing regimes in a cavity QED quantum simulator}
\end{center}
\vspace{1em}
\setcounter{section}{0}
\setcounter{subsection}{0}
\setcounter{figure}{0}
\setcounter{table}{0}
\setcounter{equation}{0}
\makeatletter
\@removefromreset{equation}{section}
\makeatother
\makeatletter
\let\@sectioncntformat\@seccntformat
\def\@hangfrom@section#1#2#3{\@hangfrom{#1#2}\MakeTextUppercase{#3}}
\makeatother
\renewcommand{\thesection}{S\arabic{section}}
\renewcommand{\thesubsection}{\Alph{subsection}}
\renewcommand{\thesubsubsection}{\arabic{subsubsection}}
\renewcommand{\thefigure}{S\arabic{figure}}
\renewcommand{\thetable}{S\arabic{table}}
\renewcommand{\theequation}{S\arabic{equation}}
\providecommand{\theHsection}{}
\renewcommand{\theHsection}{supp.\arabic{section}}
\providecommand{\theHsubsection}{}
\renewcommand{\theHsubsection}{supp.\arabic{section}.\arabic{subsection}}
\providecommand{\theHsubsubsection}{}
\renewcommand{\theHsubsubsection}{supp.\arabic{section}.\arabic{subsection}.\arabic{subsubsection}}
\providecommand{\theHfigure}{}
\renewcommand{\theHfigure}{supp.\arabic{figure}}
\providecommand{\theHtable}{}
\renewcommand{\theHtable}{supp.\arabic{table}}
\providecommand{\theHequation}{}
\renewcommand{\theHequation}{supp.\arabic{equation}}

\section{Derivation of effective spin model in the dressed basis}
In this section, we derive the effective Hamiltonian for the system starting from the full atom-cavity description in the laboratory frame. The total Hamiltonian can be written as $\hat{H}_{\mathrm{lab}}=\hat{H}_{{\mathrm{atom}}}+\hat{H}_{{\mathrm{cav}}}+\hat{H}_{{\mathrm{int}}}$, with the following second-quantized form~\cite{ThompsonMomentumExchange2024,luo2024hamiltonian,luo2025realization}:
\begin{equation}
\begin{aligned}
\hat{H}_{\rm cav} /\hbar &= \omega_c \hat{a}^\dagger \hat{a} + \epsilon( \hat{a}^\dagger e^{-i\omega_d t} + \hat{a} e^{i\omega_d t}) \\
\hat{H}_{\rm atom} /\hbar &= - \omega_{\rm hf}  \int \hat{\psi}_{\Downarrow}^{\dagger}(Z) \hat{\psi}_{\Downarrow}(Z) dZ +  \omega_{a} \int \hat{\psi}_{e}^{\dagger}(Z) \hat{\psi}_{e}(Z) dZ +  \int \frac{\Omega(Z)}{2} \hat{\psi}_{\Uparrow}^{\dagger}(Z) \hat{\psi}_{\Downarrow}(Z) dZ + \rm h.c. \\
\hat{H}_{{\mathrm{int}}} / \hbar & = \sum_{\beta_e,\beta_g} g_{\beta_e,\beta_g} \int \cos \left(kZ\right)\left[\hat{a}\hat{\psi}_{\beta_e}^{\dagger}(Z)\hat{\psi}_{\beta_g}(Z)+\hat{a}^{\dagger}\hat{\psi}_{\beta_g}^{\dagger}(Z)\hat{\psi}_{\beta_e}(Z)\right] dZ.
\end{aligned}
\end{equation}
Here, the bare cavity supports a single mode with resonance frequency $\omega_c$ and wave vector $k$, and is driven by a $\sigma_+$-polarized pump laser with amplitude $\epsilon$ at frequency $\omega_d$.  The quantization axis is taken along the cavity axis. The operator $\hat{a}^\dagger$ creates a cavity photon, and $\hat{\psi}_\beta^\dagger(Z)$ creates an atom at position $Z$ in internal state $\beta$. 
We encode an effective spin-$1/2$ degree of freedom in two long-lived hyperfine ground states of $^{87}\mathrm{Rb}$, denoted by $\beta_g=\{\ket{\Uparrow},\ket{\Downarrow}\}$, where $\ket{\Downarrow}\equiv\ket{F=1,m_F=0}$ and $\ket{\Uparrow}\equiv\ket{F=2,m_F=0}$ in the $5^2S_{1/2}$ manifold. These two states are separated by the hyperfine splitting $\omega_{\rm hf}=2\pi\times6.834$ GHz.
The ground states are coupled to an excited-state manifold $5^2P_{3/2}$ via the cavity field. The relevant excited states are $\beta_e=\{\ket{F'=3,m_{F'}=1}, \ket{F'=2,m_{F'}=1}, \ket{F'=1,m_{F'}=1}\}$, where the optical transition from $\ket{\Uparrow}$ has frequency $\omega_a$ and the excited-state manifold has natural linewidth $\Gamma$. The coupling strengths $g_{\beta_e,\beta_g}$ depend on the corresponding Clebsch-Gordan (CG) coefficients for each transition.
For simplicity, we neglect the hyperfine splitting within the excited-state manifold and treat it as effectively degenerate. We define the pump-atom detuning as $\Delta_{a}=\omega_d-\omega_a$.
In addition, a microwave field drives coherent transitions between $\ket{\Uparrow}$ and $\ket{\Downarrow}$. This drive is characterized by a mean Rabi frequency $\bar{\Omega}=2\pi\times10$ kHz and an inhomogeneous broadening with standard deviation $\sigma_\Omega=2\pi\times0.8$ kHz.

We now transform to a rotating frame defined by $\hat{H}_r/\hbar= \omega_d \hat{a}^\dagger\hat{a}+\omega_d \int \hat{\psi}^\dagger_e(Z) \hat{\psi}_e(Z)dZ$. 
In this rotating frame, we adiabatically eliminate the excited states $\beta_e$ when $|\Delta_a|\gg g_{\beta_e,\beta_g}\sqrt{\langle\hat a^\dagger\hat a\rangle},\Gamma$, so that their population remains negligible. The resulting dispersive Hamiltonian for the ground-state manifold is
\begin{equation}
\begin{aligned}
\hat{H}_{\rm disp} /\hbar &= \int \left[\frac{2 g^2}{3 \Delta_a} \hat{\psi}_{\Uparrow}^\dagger (Z) \hat{\psi}_{\Uparrow} (Z) + \frac{2 g^2}{3 (\Delta_a-\omega_{\rm hf})} \hat{\psi}_{\Downarrow}^\dagger (Z) \hat{\psi}_{\Downarrow} (Z)\right] \hat{a}^\dagger \hat{a} \cos^2(kZ) dZ \\
&+\int \frac{\Omega(Z)}{2}\left[\hat{\psi}_{\Uparrow}^{\dagger}(Z)\hat{\psi}_{\Downarrow}(Z)+\hat{\psi}_{\Downarrow}^{\dagger}(Z)\hat{\psi}_{\Uparrow}(Z)\right]dZ  - (\omega_d-\omega_c) \hat{a}^\dagger \hat{a} + \epsilon(\hat{a}^\dagger + \hat{a}).
\end{aligned}
\end{equation}
Here, $g$ denotes the atom-cavity coupling strength for the cycling transition, and the numerical prefactors arise from the relevant CG coefficients associated with the multilevel structure of the excited-state manifold.

The cavity intensity profile can be expanded as $\cos^{2}(kZ)=\tfrac{1}{4}(e^{-i2kZ}+e^{i2kZ}+2)$. The oscillatory terms describe Doppler-sensitive processes with nonzero momentum transfer, whereas the constant term describes the Doppler-insensitive process. As the atoms fall along the cavity axis, the Doppler-sensitive terms average to zero. Retaining the constant term therefore gives an effectively homogeneous atom--cavity coupling with zero net momentum transfer in the internal-state encoding.
We further introduce the pseudospin-1/2 operators in the position basis: 
\begin{equation}
\hat{j}_{x}(Z) = \frac{\hat{\psi}_{\Uparrow}^{\dagger}(Z) \hat{\psi}_{\Downarrow}(Z) + \hat{\psi}_{\Downarrow}^{\dagger}(Z) \hat{\psi}_{\Uparrow}(Z)}{2} \quad \hat{j}_{y}(Z) = \frac{\hat{\psi}_{\Uparrow}^{\dagger}(Z) \hat{\psi}_{\Downarrow}(Z) - \hat{\psi}_{\Downarrow}^{\dagger}(Z) \hat{\psi}_{\Uparrow}(Z)}{2i} \quad \hat{j}_{z}(Z) = \frac{\hat{\psi}_{\Uparrow}^{\dagger}(Z) \hat{\psi}_{\Uparrow}(Z) - \hat{\psi}_{\Downarrow}^{\dagger}(Z) \hat{\psi}_{\Downarrow}(Z)}{2}.
\end{equation} 
For simplicity, we denote $\hat{j}_{\alpha}(Z_i) \equiv \hat{j}_{\alpha,i}$ for atom $i$ at position $Z_i$, and define the collective operators $\hat{J}_{\alpha} = \sum_i \hat{j}_{\alpha,i}$ for $\alpha = x, y, z$.

The dispersive coupling Hamiltonian then simplifies to
\begin{equation}
\hat{H}_{\rm disp}/\hbar
=\tilde g_{\rm eff}\hat a^\dagger \hat a \hat J_z
+\bar\Omega \hat J_x
+\sum_i \delta\Omega_i \hat j_{x,i}
+\epsilon(\hat a^\dagger+\hat a)
-\Delta_d \hat a^\dagger \hat a,
\end{equation}
where $\Delta_d=\omega_d-\omega_c-\frac{g^2N}{6\Delta_a}-\frac{g^2N}{6(\Delta_a-\omega_{\rm hf})}$ is the dressed pump--cavity detuning and $\tilde g_{\rm eff}=\frac{g^2}{3}\left(\frac{1}{\Delta_a}-\frac{1}{\Delta_a-\omega_{\rm hf}}\right)$. The Rabi frequency experienced by atom $i$ is written as $\Omega_i=\bar\Omega+\delta\Omega_i$.
The jump operators include cavity decay at rate $\kappa$ as $\hat{L}_c = \sqrt{\kappa} \hat{a}$, and the single-particle Raman scattering processes for atom $i$:
\begin{equation}
\hat{L}_{\Uparrow\Uparrow,i}=\sqrt{\gamma_{\Uparrow\Uparrow}}\hat\sigma_{\Uparrow\Uparrow,i}\quad \hat{L}_{\Downarrow\Downarrow,i}=\sqrt{\gamma_{\Downarrow\Downarrow}}\hat\sigma_{\Downarrow\Downarrow,i}\quad \hat{L}_{\Uparrow\Downarrow,i}=\sqrt{\gamma_{\Uparrow\Downarrow}}\hat\sigma_{\Uparrow\Downarrow,i}\quad \hat{L}_{\Downarrow\Uparrow,i}=\sqrt{\gamma_{\Downarrow\Uparrow}}\hat\sigma_{\Downarrow\Uparrow,i}
\end{equation}
where $\hat \sigma_{\beta\beta',i}=\ket{\beta}_i\bra{\beta'}$ and the rates are
\begin{equation}
\gamma_{\Uparrow\Uparrow}\approx\frac{1}{9}\Gamma\langle \hat{a}^\dagger \hat{a}\rangle\frac{g^{2}}{\Delta_{a}^{2}}\quad\gamma_{\Downarrow\Downarrow}\approx\frac{17}{144}\Gamma\langle \hat{a}^\dagger \hat{a}\rangle\frac{g^{2}}{(\Delta_{a}-\omega_{\rm hf})^{2}}\quad\gamma_{\Uparrow\Downarrow}\approx\frac{5}{144}\Gamma\langle \hat{a}^\dagger \hat{a}\rangle\frac{g^{2}}{(\Delta_{a}-\omega_{\rm hf})^{2}}\quad\gamma_{\Downarrow\Uparrow}\approx\frac{5}{144}\Gamma\langle \hat{a}^\dagger \hat{a}\rangle\frac{g^{2}}{\Delta_{a}^{2}}.
\end{equation}
We also account for scattering out of the ground two-level manifold into other Zeeman sublevels:
\begin{equation}
\hat{L}_{l\Uparrow,i}=\sqrt{\gamma_{l\Uparrow}}\hat{l}_{\Uparrow,i}\quad \hat{L}_{l\Downarrow,i}=\sqrt{\gamma_{l\Downarrow}}\hat{l}_{\Downarrow,i}
\end{equation}
where $\hat{l}_{\beta_g,i}$ represents the loss of atom $i$ in $\beta_g$ state, with rates
\begin{equation}
\gamma_{l\Uparrow}=0.19\Gamma\langle \hat{a}^\dagger \hat{a}\rangle\frac{g^{2}}{\Delta_{a}^{2}}\quad\gamma_{l\Downarrow}=0.18\Gamma\langle \hat{a}^\dagger \hat{a}\rangle\frac{g^{2}}{(\Delta_{a}-\omega_{\rm hf})^{2}}.
\end{equation}
Here $\langle \hat{a}^\dagger \hat{a}\rangle$ represents the intracavity photon number. 

To diagonalize the microwave drive, we introduce the dressed basis defined by $\hat s_{x,i}=-\hat j_{z,i}$, $\hat s_{y,i}=\hat j_{y,i}$, and $\hat s_{z,i}=\hat j_{x,i}$. The dispersive Hamiltonian then becomes
\begin{equation}
\hat{H}_{\rm disp}/\hbar=-{\tilde g}_{\rm eff}\hat{a}^{\dagger}\hat{a}\hat{S}_{x} + \bar{\Omega}\hat{S}_{z}+\sum_{i}\delta \Omega_{i}\hat{s}_{z,i}+\epsilon(\hat{a}+\hat{a}^{\dagger})-\Delta_{d}\hat{a}^{\dagger}\hat{a}.
\end{equation}
We further define $2\Delta\Omega_i=\delta\Omega_i-2\tilde\mu$ by moving to a frame rotating at $2\tilde\mu\simeq-2\sigma_\Omega$. This choice makes the populated dressed-state energies positive and identifies $\Delta\Omega_i$ with the single-particle dispersion in the BCS model. Below, $\Delta\Omega_i$ denotes the shifted dispersion, while $\delta\Omega_i$ denotes the physical Rabi detuning.

\subsection{Effective spin model}
Following a similar procedure to that in Ref.~\cite{ThompsonMomentumExchange2024}, we decompose the cavity field operator into a classical steady-state amplitude and quantum fluctuations, $\hat{a} = \beta  + \hat{b}$, where $\beta = \frac{\epsilon}{\Delta_d + i\kappa/2}\sim  \frac{\epsilon}{\Delta_d }$. The operator $\hat{b}$ captures the residual quantum dynamics of the cavity field around this mean value. Substituting this decomposition into the Hamiltonian, we obtain:
\begin{equation}
\hat{H}_{\rm {disp}}'/\hbar = -
 ( g_{\rm eff}\hat{b}^{\dagger} +g_{\rm eff}^*\hat{b})\hat{S}_{x} + \bar{\Omega} \hat{S}_z  + 2\sum_j \Delta \Omega_j \hat{s}_{z,j} -\Delta_d \hat{b}^{\dagger} \hat{b}. 
\label{eqn:atomcavity}
\end{equation}
where 
$g_{\rm eff}=\tilde{g}_{\rm eff}|\beta|$. In the regime $\sqrt{N}g_{\rm eff} \ll |\Delta_d|$, the cavity fluctuations $\hat{b}$ can be adiabatically eliminated. Applying standard second-order perturbation theory yields the following effective spin-exchange Hamiltonian in a frame rotating about the dressed-basis $z$ axis at angular frequency $\bar\Omega$:
\begin{equation}
\hat{H}_{{\rm eff}}/\hbar=\frac{g_{\rm eff}^{2}}{4}\left[\frac{\Delta_d- \bar{\Omega}}{(\Delta_d- \bar{\Omega})^{2}+(\kappa/2)^{2}}\hat{S}_{-}\hat{S}_{+}+\frac{\Delta_d+ \bar{\Omega}}{(\Delta_d+ \bar{\Omega})^{2}+(\kappa/2)^{2}}\hat{S}_{+}\hat{S}_{-}\right] + 2\sum_j \Delta \Omega_j \hat{s}_{z,j},
\end{equation}
together with collective dissipative processes described by the jump operators
\begin{equation}
\hat{L}_{{\rm eff},+}=\sqrt{\frac{g_{\rm eff}^{2}}{4}\frac{\kappa}{(\Delta_d- \bar{\Omega})^{2}+(\kappa/2)^{2}}}\hat{S}_{+} \quad \hat{L}_{{\rm eff},-}=\sqrt{\frac{g_{\rm eff}^{2}}{4}\frac{\kappa}{(\Delta_d+ \bar{\Omega})^{2}+(\kappa/2)^{2}}}\hat{S}_{-}.
\end{equation}

In the experimental regime, $\bar{\Omega}=2\pi\times10$ kHz and $\kappa=2\pi\times56$ kHz are both much smaller than the detuning magnitude $|\Delta_d|=2\pi\times1.9$ MHz. Under this condition, the effective Hamiltonian simplifies to:
\begin{equation}
\hat{H}_{\rm eff}/\hbar \approx - \chi \hat{S}_+ \hat{S}_- + 2\sum_j \Delta \Omega_j \hat{s}_{z,j},
\label{eq:spin}
\end{equation}
where the exchange coupling is given by $\chi = -\frac{g_{\rm eff}^2}{2\Delta_d}$. For typical experimental parameters with total atom number $N \approx 8.8 \times 10^4$ and $|\beta|^2 \approx 4\times10^3$, we obtain $\chi N \approx 2\pi \times 4.3$ kHz, in good agreement with experimental observations.

\subsection{Atom-cavity simulation}
However, as shown in Fig.~2 of the main text, the cavity field exhibits a pronounced time-dependent oscillation arising from its dispersive coupling to the driven collective spin. The resulting amplitude of the cavity-resonance-frequency modulation is on the order of $\tilde{g}_{\rm eff} N \approx 2\pi \times 550$ kHz, which is no longer negligible compared to the detuning magnitude $|\Delta_d|$. In this regime, the adiabatic elimination of the cavity field is not quantitatively accurate. To capture the dynamics more faithfully, we instead retain the cavity field as an active degree of freedom and consider the following mean-field equations of motion:
\begin{equation}
\begin{aligned}
\frac{d}{dt}\left\langle \hat{a}\right\rangle &=i \tilde{g}_{\rm eff}\left\langle \hat{a}\right\rangle \left\langle \hat{S}_{x}\right\rangle -i\epsilon+i\Delta_d\left\langle \hat{a}\right\rangle - \frac{\kappa}{2} \left\langle \hat{a}\right\rangle \\
\frac{d}{dt}\left\langle \hat{s}_{x,j}\right\rangle &=-(\bar{\Omega}+2\Delta \Omega_j)\left\langle \hat{s}_{y,j}\right\rangle -(\gamma_{\Downarrow\Uparrow}+\gamma_{\Uparrow\Downarrow}+\frac{\gamma_{l\Uparrow}+\gamma_{l\Downarrow}}{2})\left\langle \hat{s}_{x,j}\right\rangle +(\frac{\gamma_{\Downarrow\Uparrow}-\gamma_{\Uparrow\Downarrow}}{2}+\frac{\gamma_{l\Uparrow}-\gamma_{l\Downarrow}}{4})\langle \hat{n}_j \rangle \\
\frac{d}{dt}\left\langle \hat{s}_{y,j}\right\rangle &=(\bar{\Omega}+2\Delta \Omega_j)\left\langle \hat{s}_{x,j}\right\rangle +\tilde{g}_{\rm eff} |\langle \hat{a} \rangle|^{2}\left\langle \hat{s}_{z,j}\right\rangle -\frac{1}{2}(\gamma_{\Uparrow\Uparrow}+\gamma_{\Downarrow\Downarrow}+\gamma_{\Downarrow\Uparrow}+\gamma_{\Uparrow\Downarrow}+\gamma_{l\Uparrow}+\gamma_{l\Downarrow})\left\langle \hat{s}_{y,j}\right\rangle \\
\frac{d}{dt}\left\langle \hat{s}_{z,j}\right\rangle &=-\tilde{g}_{\rm eff}|\langle \hat{a} \rangle|^{2}\left\langle \hat{s}_{y,j}\right\rangle -\frac{1}{2}(\gamma_{\Uparrow\Uparrow}+\gamma_{\Downarrow\Downarrow}+\gamma_{\Downarrow\Uparrow}+\gamma_{\Uparrow\Downarrow}+\gamma_{l\Uparrow}+\gamma_{l\Downarrow})\left\langle \hat{s}_{z,j}\right\rangle \\
\frac{d}{dt}\langle \hat n_j\rangle
&=-\frac{\gamma_{l\Uparrow}+\gamma_{l\Downarrow}}{2}\langle \hat n_j\rangle
+(\gamma_{l\Uparrow}-\gamma_{l\Downarrow})
\langle \hat{s}_{x,j}\rangle.
\label{eq:ac}
\end{aligned}
\end{equation}
Here $\langle \hat{n}_j \rangle$ is the  population in the two level system.  We note that the single-particle dissipation rates implicitly depend on the intracavity photon number $|\langle \hat{a} \rangle|^{2}$. The theoretical results shown in Figs.~3--5 of the main text are obtained by numerically integrating Eq.~\eqref{eq:ac}.

To understand the single-particle dissipation process in the system, we consider the single-particle limit by neglecting atom-light interactions and inhomogeneity, such that $\tilde g_{\rm eff}=0$ and $\Delta\Omega_j=0$.
We define $\Gamma_\perp\equiv(\gamma_{\Uparrow\Uparrow}+\gamma_{\Downarrow\Downarrow}+\gamma_{\Uparrow\Downarrow}+\gamma_{\Downarrow\Uparrow}+\gamma_{l\Uparrow}+\gamma_{l\Downarrow})/2$, $\Gamma_n\equiv(\gamma_{l\Uparrow}+\gamma_{l\Downarrow})/2$, and $\Gamma_x=\gamma_{\Uparrow\Downarrow}+\gamma_{\Downarrow\Uparrow}+\Gamma_n$. The single-particle equations of motion can then be written as
\begin{equation}
\begin{aligned}
\frac{d}{dt}\langle\hat s_{x,j}\rangle
&=-\bar\Omega\langle\hat s_{y,j}\rangle-
\Gamma_x \langle\hat s_{x,j}\rangle +\left(
\frac{\gamma_{\Downarrow\Uparrow}-\gamma_{\Uparrow\Downarrow}}{2}
+\frac{\gamma_{l\Uparrow}-\gamma_{l\Downarrow}}{4}
\right)\langle\hat n_j\rangle,\\
\frac{d}{dt}\langle\hat s_{y,j}\rangle
&=\bar\Omega\langle\hat s_{x,j}\rangle-\Gamma_\perp\langle\hat s_{y,j}\rangle,\quad \frac{d}{dt}\langle\hat s_{z,j}\rangle
=-\Gamma_\perp\langle\hat s_{z,j}\rangle,\\
\frac{d}{dt}\langle\hat n_j\rangle
&=-\Gamma_n\langle\hat n_j\rangle
+(\gamma_{l\Uparrow}-\gamma_{l\Downarrow})\langle\hat s_{x,j}\rangle .
\end{aligned}
\end{equation}

The longitudinal spin component is decoupled from the other variables and therefore decays exponentially as
$\langle\hat s_{z,j}(t)\rangle=\langle\hat s_{z,j}(0)\rangle e^{-\Gamma_\perp t}$.
To characterize the decay of the transverse spin contrast, we define $C\equiv\sqrt{\langle\hat s_{x,j}\rangle^2+\langle\hat s_{y,j}\rangle^2}$, which satisfies
\begin{equation}
\frac{dC^2}{dt} =-2\Gamma_x \langle\hat s_{x,j}\rangle^2 -2\Gamma_\perp\langle\hat s_{y,j}\rangle^2+2\left( \frac{\gamma_{\Downarrow\Uparrow}-\gamma_{\Uparrow\Downarrow}}{2} +\frac{\gamma_{l\Uparrow}-\gamma_{l\Downarrow}}{4} \right) \langle\hat n_j\rangle\langle\hat s_{x,j}\rangle .
\end{equation}
In the strong-drive regime, $\bar\Omega$ is much larger than all single-particle dissipation rates. The transverse spin components therefore undergo rapid oscillations compared with the dissipative dynamics. Averaging over these fast oscillations, we obtain
$\overline{\langle\hat s_{x,j}\rangle^2}
\simeq
\overline{\langle\hat s_{y,j}\rangle^2}
\simeq C^2/2$
and
$\overline{\langle\hat s_{x,j}\rangle}\simeq0$.
As a result, the slowly varying transverse contrast satisfies
\begin{equation}
\frac{dC^2}{dt} \simeq -\left(\Gamma_x + \Gamma_\perp \right)C^2 \Rightarrow C(t)\simeq C(0)e^{-\left(\Gamma_x + \Gamma_\perp \right) t/2}
\end{equation}


The effective decay rate of the transverse contrast is therefore
$\Gamma_{xy}\equiv(\Gamma_x+\Gamma_\perp)/2$, while the longitudinal-spin
decay rate is $\Gamma_z\equiv\Gamma_\perp$. Their ratio is
\begin{equation}
    \frac{\Gamma_{xy}}{\Gamma_z}
    =
    \frac{\Gamma_x+\Gamma_\perp}{2\Gamma_\perp}
    \simeq 0.89
\end{equation}
for the experimental parameters discussed above. Thus, the transverse and
longitudinal decay rates differ by only approximately $11\%$. Consequently,
the slowly varying total Bloch-vector length is well approximated by a single
exponential envelope, $e^{-t/\tau}$, over the experimental time window. We
therefore use a common effective decay constant $\tau$ in the fitting model.

\section{Analytical insights for the spin model \label{sec:spin}}
\subsection{Prepared initial spin texture state}
In our simulator, the inhomogeneous detuning $\Delta\Omega_j$ plays the role of the single-particle dispersion. The prepared spin texture used throughout this work is
\begin{equation}
\vec{\mathbf s}_j^{(0)}
=\frac{1}{2}\left(
\cos(\theta_0+\delta\Omega_jT_d),\,
0,\,
-\sin(\theta_0+\delta\Omega_jT_d)
\right),
\label{eq:fgs}
\end{equation}
where $\delta\Omega_j=2(\Delta\Omega_j+\tilde\mu)$, $\theta_0$ is the central tipping angle measured from the equator, and $\sigma_\theta=\sigma_\Omega T_d$ is the rms angular spread.

\begin{figure}[t]
\centering
\includegraphics[width=0.7\linewidth]{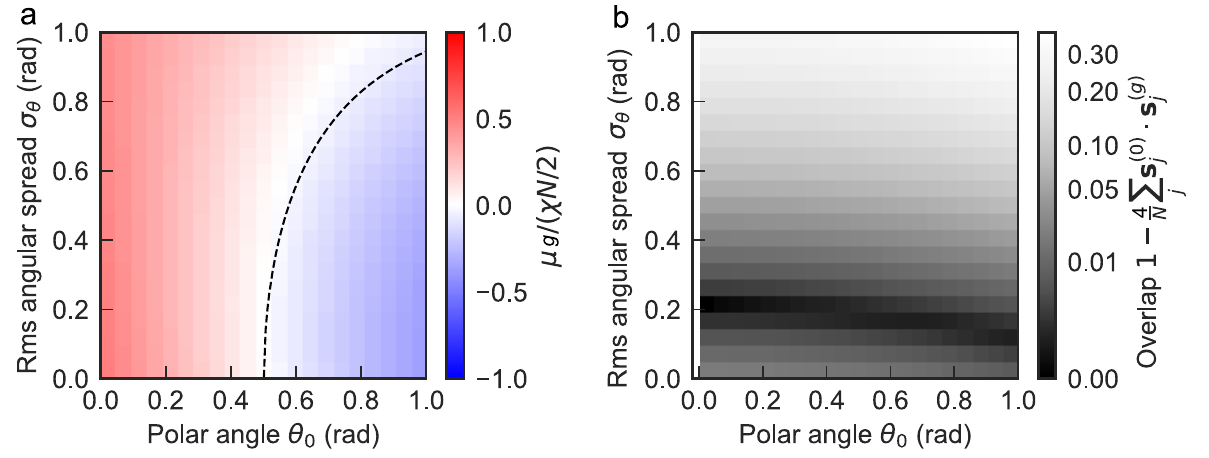}
\caption{\textbf{Self-consistent ground states of the cavity simulator.} (a) Ground-state chemical potential $\mu_g$, normalized by $\chi N/2$, as a function of the initial tipping angle $\theta_0$ and rms angular spread $\sigma_\theta=\sigma_\Omega T_d$. The dashed contour at $\mu_g=0$ separates the BCS-like ($\mu_g>0$) and BEC-like ($\mu_g<0$) regions. (b) Mismatch $1-\frac{4}{N}\sum_j\vec{\mathbf s}_j^{(0)}\cdot\vec{\mathbf s}_j^{(g)}$ between the prepared spin texture $\vec{\mathbf s}_j^{(0)}$ and its corresponding self-consistent ground state $\vec{\mathbf s}_j^{(g)}$.}
\label{fig:s1}
\end{figure}

To locate these prepared states relative to the equilibrium BCS--BEC crossover, we calculate the corresponding self-consistent mean-field ground state of our simulator. The standard ground-state construction applies after replacing the single-particle kinetic energy by the shifted microwave dispersion $\Delta\Omega_j$. In a frame rotating at $2\mu_g$, the effective field and the aligned ground-state spin are
\begin{equation}
\mathbf B_j^{(g)}=(\Delta_g,0,\mu_g-\Delta\Omega_j),\qquad
\vec{\mathbf s}_j^{(g)}=\frac{\mathbf B_j^{(g)}}{2E_j^{(g)}},\qquad
E_j^{(g)}=\sqrt{\Delta_g^2+(\mu_g-\Delta\Omega_j)^2}.
\label{eq:simulator-gs}
\end{equation}
For each prepared texture specified by $(\theta_0,\sigma_\theta)$, we determine $\mu_g$ and $\Delta_g$ from the number equation $\sum_j s_{j,z}^{(g)}=\sum_j s_{j,z}^{(0)}$ and gap equation $\Delta_g=\chi\sum_j s_{j,x}^{(g)}$.
Fig.~\ref{fig:s1}a shows the resulting ground-state chemical potential $\mu_g$. The dashed contour at $\mu_g=0$ separates the BCS-like region with $\mu_g>0$ from the BEC-like region with $\mu_g<0$.

We quantify how closely the experimentally prepared texture follows this ground state using the mismatch in the normalized overlap $1-\frac{4}{N}\sum_j\vec{\mathbf s}_j^{(0)}\cdot\vec{\mathbf s}_j^{(g)}$, shown in Fig.~\ref{fig:s1}b. A nonzero mismatch means that the prepared texture is not an eigenstate of the post-quench Hamiltonian and therefore initiates the nonequilibrium dynamics studied below.

\subsection{Approximate solution for the decay of the Higgs oscillations~\label{sec:approx}}

In this section, we estimate the decay of the Higgs oscillations in different parameter regimes. The long-time dynamics are characterized by the asymptotic parameters $\mu_\infty$ and $\Delta_\infty$. We first separate the exact precession in the asymptotic field from the residual self-consistent response. We then show that the residual field produces a $t^{-1/2}$ power-law decay in the BCS-like regime with $\mu_\infty>0$ and an exponentially damped response in the polarized BEC-like limit.

In the continuum limit, we denote the shifted single-particle dispersion $\Delta\Omega_j$ by $\Omega$ and use the normalized Gaussian density
\[
g(\Omega)=\sqrt{\frac{2}{\pi}}\frac{1}{\sigma_\Omega}\exp\!\left[-\frac{2(\Omega+\tilde\mu)^2}{\sigma_\Omega^2}\right],
\qquad \int d\Omega\,g(\Omega)=1.
\]
With this convention, $\delta\Omega$ has zero mean and rms width $\sigma_\Omega$. We work in the rotating frame where the long-time pairing field is real. The time-dependent pairing field is defined self-consistently by
\begin{equation}
\Delta(t)\equiv\Delta_x(t)-i\Delta_y(t)=\chi N\int d\Omega\,g(\Omega)s_{\Omega,-}(t),\qquad s_{\Omega,-}(t)\equiv s_{\Omega,x}(t)-is_{\Omega,y}(t).
\label{eq:dynamic-gap}
\end{equation}
For a spin labeled by the continuous detuning variable $\Omega$, we define
\begin{equation}
\epsilon_\Omega=\Omega-\mu_\infty,\qquad
{\bf B}_{\Omega}^{\infty}=(\Delta_\infty,0,-\epsilon_\Omega),\qquad{\bf B}_{\Omega}^{\perp}=(\epsilon_\Omega,0,\Delta_\infty),\quad 
E_\Omega=|{\bf B}_{\Omega}^{\infty}|=\sqrt{\Delta_\infty^2+\epsilon_\Omega^2}.
\end{equation}
Here ${\bf B}_{\Omega}^{\infty}$ represents the frozen asymptotic field about which the spin precesses.
The initial spin is
\begin{equation}
\vec{\mathbf s}_\Omega(0)=\frac{1}{2}\bigl(\cos\Theta_\Omega,0,-\sin\Theta_\Omega\bigr),\qquad
\Theta_\Omega=\theta_0+\delta\Omega T_d,
\end{equation}
with $\delta \Omega=2(\Omega+\tilde \mu)$.
We decompose the initial spin into components parallel and perpendicular to the asymptotic field,
\begin{equation}
\vec{\mathbf s}_\Omega(0)=\vec{\mathbf s}_{\Omega}^{\parallel}+\vec{\mathbf s}_{\Omega}^{\perp}(0),\qquad
\vec{\mathbf s}_{\Omega}^{\parallel}=\frac{P_\Omega}{E_\Omega^2}{\bf B}_{\Omega}^{\infty}, \qquad
\vec{\mathbf s}_{\Omega}^{\perp}(0)=\frac{Q_\Omega}{E_\Omega^2}{\bf B}_{\Omega}^{\perp}
\end{equation}
with
\begin{eqnarray}
P_\Omega
&=&
\mathbf{B}_{\Omega}^{\infty}\cdot\vec{\mathbf s}_\Omega(0)
=
\frac{1}{2}
\left(
\Delta_\infty\cos\Theta_\Omega
+
\epsilon_\Omega\sin\Theta_\Omega
\right),
\\
Q_\Omega
&=&
\mathbf{B}_{\Omega}^{\perp}\cdot\vec{\mathbf s}_\Omega(0)
=
\frac{1}{2}
\left(\epsilon_\Omega\cos\Theta_\Omega
-\Delta_\infty\sin\Theta_\Omega
\right).
\end{eqnarray}
The stationary component $\vec{\mathbf s}_{\Omega}^{\parallel}$ is parallel to ${\bf B}_{\Omega}^{\infty}$ and therefore does not precess.
In contrast, the perpendicular component precesses around ${\bf B}_{\Omega}^{\infty}$ and produces the Higgs oscillation.

If we denote as $\vec{\mathbf s}_{\Omega}^{(0)}(t)$ the single-particle solution generated by the frozen asymptotic field ${\bf B}_{\Omega}^{\infty}$,
\begin{equation}
\dot{\vec{\mathbf s}}_{\Omega}^{(0)}=-2{\bf B}_{\Omega}^{\infty} \times \vec{\mathbf s}^{(0)}_{\Omega}, \qquad
\vec{\mathbf s}_{\Omega}^{(0)}(0)= \vec{\mathbf s}_\Omega(0),
\end{equation} 
we find
\begin{equation}
\begin{aligned}
s_{\Omega,x}^{(0)}(t)&=\frac{\Delta_\infty P_\Omega}{E_\Omega^2}
+\frac{\epsilon_\Omega Q_\Omega}{E_\Omega^2}\cos(2E_\Omega t),\\
s_{\Omega,y}^{(0)}(t)&=\frac{Q_\Omega}{E_\Omega}\sin(2E_\Omega t),\\
s_{\Omega,z}^{(0)}(t)&=-\frac{\epsilon_\Omega P_\Omega}{E_\Omega^2}
+\frac{\Delta_\infty Q_\Omega}{E_\Omega^2}\cos(2E_\Omega t).
\end{aligned}
\label{eq:frozen-spin}
\end{equation}
Therefore  $s_{\Omega,-}^{(0)}(t)$ can be written as
\begin{equation}
\begin{aligned}
s_{\Omega,-}^{(0)}(t)&=s_{\Omega,x}^{\parallel}
+a_+(\Omega)e^{-2iE_\Omega t}+a_-(\Omega)e^{2iE_\Omega t}\\
a_\pm(\Omega)&=\frac{\epsilon_\Omega\pm E_\Omega}{4E_\Omega^2}
\left(\epsilon_\Omega\cos\Theta_\Omega-\Delta_\infty\sin\Theta_\Omega\right),
\end{aligned}
\end{equation}
The first term represents the static $x$ contribution from $\vec{\mathbf s}_{\Omega}^{\parallel}$, while the second and third terms represent the time-dependent contribution from $\vec{\mathbf s}_{\Omega}^{\perp,(0)}(t)$.

From the gap equation, $
\chi N\int d\Omega\,g(\Omega)s_{\Omega,x}^{\parallel}=\Delta_\infty $ and  particle  number conservation $\int d\Omega g(\Omega)s_{\Omega,z}^{(0)}(t)= \int d\Omega\,g(\Omega)s_{\Omega,z}^{(0)}(0)$, and after  introducing $u_+\equiv \mu_\infty+i\Delta_\infty$ so we can write  
\begin{equation}
{\rm Im}\left[\frac{e^{i\Theta_\Omega}}{2\Omega-2u_+}\right]
=\frac{P_\Omega}{E_\Omega^2}
\quad {\rm Re}\left[\frac{e^{i\Theta_\Omega}}{2\Omega-2u_+}\right]
=\frac{Q_\Omega}{E_\Omega^2},
\end{equation} 
we obtain, 
\begin{equation}
\int d\Omega g(\Omega)
\mathrm{Im} \Big[\frac{e^{i\Theta_\Omega}}{2\Omega-2u_+}\Big]=\frac{1}{\chi N},  \quad \int d\Omega g(\Omega)
\mathrm{Re} \Big[\frac{e^{i\Theta_\Omega}}{\Omega-u_+}\Big]=0.
\label{eq:imag}
\end{equation}

These two conditions correspond to the imaginary and real parts of the equation obtained from the full Lax formalism, providing an intuitive geometric interpretation of the integrability-based result in Sec.~\ref{sec:lax}.

The frozen-field trajectories are not self-consistent on their own because their oscillating transverse components generate a residual collective field. We therefore decompose the spin and field as
\begin{equation}
\vec{\mathbf s}_\Omega(t)=\vec{\mathbf s}_\Omega^{(0)}(t)+\delta\vec{\mathbf s}_\Omega(t),\qquad
\mathbf B_\Omega(t)=\mathbf B_\Omega^{\infty}+\delta\mathbf B(t).
\end{equation}
Because the pairing field is generated collectively by the transverse spin components, the spin decomposition directly gives
\begin{equation}
\begin{aligned}
\delta\Delta(t)
&\equiv \Delta(t)-\Delta_\infty=\chi N\int d\Omega\,g(\Omega)
\left[s_{\Omega,-}^{(0)}(t)-s_{\Omega,-}^{\parallel}
+\delta s_{\Omega,-}(t)\right]\\
&=\delta\Delta^{(0)}(t)
+\chi N\int d\Omega\,g(\Omega)\delta s_{\Omega,-}(t),
\end{aligned}
\label{eq:feedback-self-consistency}
\end{equation}
where we define
\begin{equation}
\begin{aligned}
\delta s_{\Omega,-}(t)&\equiv x_\Omega(t)-iy_\Omega(t),\\
\delta\Delta^{(0)}(t)
&\equiv\chi N\int d\Omega\,g(\Omega)
\left[s_{\Omega,-}^{(0)}(t)-s_{\Omega,-}^{\parallel}\right]
\equiv\delta\Delta_x^{(0)}(t)-i\delta\Delta_y^{(0)}(t).
\end{aligned}
\label{eq:frozen-source}
\end{equation}
Here $\delta\Delta^{(0)}(t)$ is the part of the residual field generated by $\vec{\mathbf s}_{\Omega}^{\perp,(0)}(t)$, whereas the second term in Eq.~\eqref{eq:feedback-self-consistency} is generated by the additional spin response $\delta\vec{\mathbf s}_\Omega(t)$.

Since $\delta\Delta(t)=\delta\Delta_x(t)-i\delta\Delta_y(t)$, the residual field acting on each spin is $\delta\mathbf B(t)=\bigl(\delta\Delta_x(t),\delta\Delta_y(t),0\bigr)$, where $\delta\Delta_x(t)=\Delta_x(t)-\Delta_\infty$ and $\delta\Delta_y(t)=\Delta_y(t)$.
We further define the residual field as:
\begin{equation}
\delta\vec{\mathbf s}_\Omega(t)
\equiv\bigl(x_\Omega(t),y_\Omega(t),z_\Omega(t)\bigr)
\end{equation}
whose linearized spin response obeys
\begin{equation}
\frac{d}{dt}\delta\vec{\mathbf s}_\Omega
=-2\mathbf B_\Omega^{\infty}\times\delta\vec{\mathbf s}_\Omega
-2\delta\mathbf B(t)\times\vec{\mathbf s}_\Omega^{(0)}(t),
\label{eq:linear-spin-response}
\end{equation}
thus leads to
\begin{equation}
\begin{aligned}
\dot{x}_\Omega&=-2\epsilon_\Omega y_\Omega
-2s_{\Omega,z}^{(0)}(t)\delta\Delta_y(t),\\
\dot{y}_\Omega&=2\epsilon_\Omega x_\Omega+2\Delta_\infty z_\Omega
+2s_{\Omega,z}^{(0)}(t)\delta\Delta_x(t),\\
\dot{z}_\Omega&=-2\Delta_\infty y_\Omega
-2s_{\Omega,y}^{(0)}(t)\delta\Delta_x(t)
+2s_{\Omega,x}^{(0)}(t)\delta\Delta_y(t).
\end{aligned}
\label{eq:xyz-response}
\end{equation}
Equations~\eqref{eq:linear-spin-response} and \eqref{eq:xyz-response} determine the linear spin response for a specified residual field. Self-consistency closes the dynamics through $\delta\Delta_x(t)=\delta\Delta_x^{(0)}(t)+\chi N\int d\Omega\,g(\Omega)x_\Omega(t)$ and $\delta\Delta_y(t)=\delta\Delta_y^{(0)}(t)+\chi N\int d\Omega\,g(\Omega)y_\Omega(t)$.

\subsubsection{BCS-like regime}
Using Eq.~\eqref{eq:frozen-source} and the mode expansion in Eq.~\eqref{eq:frozen-spin}, the frozen-field source becomes
\begin{equation}
\begin{aligned}
\delta \Delta^{(0)}(t)
&=\chi N\int d\Omega\,g(\Omega)\left[a_+(\Omega)e^{-2iE_\Omega t}+a_-(\Omega)e^{2iE_\Omega t}\right]\\
&=\frac{\chi N}{8}\sum_{\nu=\pm1}\sum_{\lambda=\pm1}e^{-i\lambda\theta_0}
\int d\Omega\,g(\Omega)
\frac{(\epsilon_\Omega+\nu E_\Omega)(\epsilon_\Omega-i\lambda\Delta_\infty)}
{E_\Omega^2}
e^{-i(2\nu E_\Omega t+\lambda\delta\Omega T_d)}.
\end{aligned}
\label{eq:frozen-integral}
\end{equation}
The long-time behavior of this integral is controlled by the saddle point at $\Omega=\mu_\infty$. For the BCS parameters considered here, this saddle point lies within the appreciable support of the Gaussian distribution and therefore carries a finite spectral weight.
Defining the shifted nonequilibrium chemical potential $\mu'_\infty\equiv\mu_\infty+\tilde\mu$ and using $\epsilon_\Omega=\Omega-\mu_\infty$, we expand around the saddle to obtain
\begin{equation}
E_\Omega=\Delta_\infty+\frac{\epsilon_\Omega^2}{2\Delta_\infty}+O(\epsilon_\Omega^4). \label{eq:expand}
\end{equation}
The saddle-point integral for Eq.~\eqref{eq:frozen-integral} then gives
\begin{equation}
\delta\Delta_y^{(0)}(t)
\simeq-\frac{C}{\sqrt{t}}
\sin\left(2\Delta_\infty t+\frac{\pi}{4}\right),\qquad
\delta\Delta_x^{(0)}(t)=O(t^{-3/2}),
\label{eq:frozen-bcs-tail}
\end{equation}
where
\begin{equation}
C=\chi N\frac{\sqrt{\Delta_\infty}}{\sqrt{2}\sigma_\Omega}
\exp\left[-\frac{2(\mu'_\infty)^2}{\sigma_\Omega^2}\right]
\sin\left(\theta_0+2\mu'_\infty T_d\right).
\end{equation}
Equation~\eqref{eq:frozen-bcs-tail} is the frozen-field dephasing result. In particular, it should not by itself be identified with the full self-consistent order-parameter response.

We next evaluate the leading feedback correction. Since $\delta\Delta_x^{(0)}(t)=O(t^{-3/2})$ is subleading, the dominant response is driven by $\delta\Delta_y^{(0)}(t)$.
We retain the dominant drive $\delta\Delta_y^{(0)}(t)$ in Eq.~\eqref{eq:xyz-response} and replace the frozen-spin coefficients multiplying this drive by their stationary components. This gives
\begin{equation}
\dot{x}_\Omega=-2\epsilon_\Omega y_\Omega-2s_{\Omega,z}^{\infty}\delta\Delta_y^{(0)}(t),\qquad
\dot{y}_\Omega=2\epsilon_\Omega x_\Omega+2\Delta_\infty z_\Omega,\qquad
\dot{z}_\Omega=-2\Delta_\infty y_\Omega+2s_{\Omega,x}^{\infty}\delta\Delta_y^{(0)}(t).
\end{equation}
Taking the time derivative shows that $d({\bf B}_{\Omega}^{\infty}\cdot\delta\vec{\mathbf s}_\Omega)/dt=0$ and for the initial condition $\delta\vec{\mathbf s}_\Omega(0)=0$, we find ${\bf B}_{\Omega}^{\infty}\cdot\delta\vec{\mathbf s}_\Omega=\Delta_\infty x_\Omega-\epsilon_\Omega z_\Omega=0$. Eliminating $y_\Omega$ and $z_\Omega$ then gives
\begin{equation}
\ddot{x}_\Omega+4E_\Omega^2x_\Omega
=-2s_{\Omega,z}^{\infty}\frac{d}{dt}\delta\Delta_y^{(0)}(t)
=2\frac{\epsilon_\Omega P_\Omega}{E_\Omega^2}\frac{d}{dt}\delta\Delta_y^{(0)}(t).
\label{eq:forced-x}
\end{equation}
Equation~\eqref{eq:forced-x} has the exact causal solution
\begin{equation}
x_\Omega(t)=\frac{\epsilon_\Omega P_\Omega}{E_\Omega^3}
\int_0^t dt'\,\sin\left[2E_\Omega(t-t')\right]
\frac{d}{dt'}\delta\Delta_y^{(0)}(t').
\label{eq:retarded-x}
\end{equation}

The derivative of the long-time tail in Eq.~\eqref{eq:frozen-bcs-tail} is
\begin{equation}
\frac{d}{dt'}\delta\Delta_y^{(0)}(t')
=-\frac{2C\Delta_\infty}{\sqrt{t'}}
\cos\!\left(2\Delta_\infty t'+\phi\right)
+O(t'^{-3/2}),
\qquad \phi=\frac{\pi}{4},
\label{eq:drive-derivative}
\end{equation}
Substituting Eq.~\eqref{eq:drive-derivative} into Eq.~\eqref{eq:retarded-x} produces two terms with frequencies $2(E_\Omega-\Delta_\infty)$ and $2(E_\Omega+\Delta_\infty)$. For the long-time behavior, we keep the first term because it becomes slowly varying near $E_\Omega=\Delta_\infty$. The second term always oscillates rapidly; it produces no resonant contributions and does not change the long-time decay exponent. We then obtain
\begin{equation}
x_\Omega(t)
\simeq-\frac{C\Delta_\infty\epsilon_\Omega P_\Omega}{E_\Omega^3}
\operatorname{Im}\!\left[
e^{i(2E_\Omega t+\phi)}
\int^{t}\frac{dt'}{\sqrt{t'}}
e^{-2i(E_\Omega-\Delta_\infty)t'}
\right].
\label{eq:resonant-exponential}
\end{equation}
For every fixed $\epsilon_\Omega\neq0$, the $t'$ integral is rapidly oscillating at long times. Integrating by parts gives
\begin{equation}
x_\Omega(t)
\simeq-\frac{C\Delta_\infty P_\Omega}
{E_\Omega^2\epsilon_\Omega}
\frac{\cos(2\Delta_\infty t+\phi)}{\sqrt{t}}.
\label{eq:outer-x}
\end{equation}
For each fixed $\Omega\neq\mu_\infty$, this response decays as $t^{-1/2}$, but its prefactor contains an apparent singularity at $\epsilon_\Omega=0$. Because the collective response requires integration through this point, we must verify that the integrated response remains finite. Moreover, the integration-by-parts result is not uniform near resonance. Using Eq.~\eqref{eq:expand}, the phase accumulated over time $t$ becomes of order unity when $\epsilon_\Omega^2t/\Delta_\infty=O(1)$. We therefore introduce the time-dependent resonance scale
\begin{equation}
\epsilon_*(t)\equiv\sqrt{\frac{\Delta_\infty}{t}},
\qquad
\begin{cases}
|\epsilon_\Omega|\gg\epsilon_*(t), & \text{nonresonant region},\\
|\epsilon_\Omega|\lesssim\epsilon_*(t), & \text{resonant region}.
\end{cases}
\label{eq:resonant-window}
\end{equation}
In the nonresonant region, Eq.~\eqref{eq:outer-x} applies. In the collective integral, the remaining factor $g(\Omega)P_\Omega/E_\Omega^2$ is finite and smooth at $\Omega=\mu_\infty$. Pairing contributions from equal and opposite detunings, the leading terms cancel because $1/\epsilon_\Omega$ is odd. The difference between the two smooth factors is $O(|\epsilon_\Omega|)$ and removes the denominator. The nonresonant collective contribution therefore remains finite and scales as $t^{-1/2}$.

In the resonant region, the time integral in Eq.~\eqref{eq:resonant-exponential} obeys the simple bound
\begin{equation}
\left|
\int^{t}\frac{dt'}{\sqrt{t'}}
e^{-2i(E_\Omega-\Delta_\infty)t'}
\right|
\leq\int^{t}\frac{dt'}{\sqrt{t'}}
=O(\sqrt{t}).
\end{equation}
Since $|\epsilon_\Omega|\lesssim\epsilon_*(t)=O(t^{-1/2})$ in this region, Eq.~\eqref{eq:resonant-exponential} gives $x_\Omega(t)=O(1)$. At exact resonance, $\epsilon_\Omega=0$ and the response vanishes. The resonant region has width $O(t^{-1/2})$, so its collective contribution is also at most $O(t^{-1/2})$.

Both regions therefore produce a collective response with no decay slower than $t^{-1/2}$. Generically, provided that the weighted contributions do not cancel, their combined leading behavior is
\begin{equation}
\chi N\int d\Omega\,g(\Omega)x_\Omega(t)
=\frac{C\mathcal K}{\sqrt{t}}
\cos(2\Delta_\infty t+\phi)+o(t^{-1/2}),
\label{eq:kappa-response}
\end{equation}
where $\mathcal K$ is a finite real coefficient determined by the complete weighted response.
Combining the frozen-field and feedback contributions gives
\begin{equation}
\delta \Delta(t)
\approx \frac{C}{\sqrt{t}}
\left[
\mathcal K\cos\left(2\Delta_\infty t+\phi\right)
+i\sin\left(2\Delta_\infty t+\phi\right)
\right],
\label{eq:delta-Sminus-bcs}
\end{equation}
where the imaginary term originates from the frozen-field oscillation of the perpendicular component, while the self-consistent feedback generates a real amplitude response with the same $t^{-1/2}$ envelope.

\subsubsection{BEC-like regime}

We now consider the polarized BEC-like limit in which $\Delta_\infty\approx0$ and the frozen-field solution satisfies $s_{\Omega,z}^{(0)}(t)\approx-1/2$, while $s_{\Omega,x}^{(0)}(t)\approx s_{\Omega,y}^{(0)}(t)\approx0$. In this limit, the $z_\Omega$ equation decouples, and Eq.~\eqref{eq:xyz-response} reduces to
\begin{equation}
\dot{x}_\Omega=-2\epsilon_\Omega y_\Omega+\delta\Delta_y(t),\qquad
\dot{y}_\Omega=2\epsilon_\Omega x_\Omega-\delta\Delta_x(t).
\end{equation}
Using $\delta s_{\Omega,-}=x_\Omega-iy_\Omega$, already defined in Eq.~\eqref{eq:frozen-source}, the two component equations combine into
\begin{equation}
\begin{aligned}
\frac{d}{dt}\delta s_{\Omega,-}(t)
&=-2i\epsilon_\Omega\delta s_{\Omega,-}(t)+i\delta\Delta(t)\\
&=-2i\epsilon_\Omega\delta s_{\Omega,-}(t)
+i\chi N\int d\Omega'\,g(\Omega')\delta s_{\Omega',-}(t).
\end{aligned}
\label{eq:bec-linear}
\end{equation}

Because $\Delta_\infty\approx 0$ and the polarized reference state has no transverse component, $\delta\Delta(t)$ is the full pairing field, while $x_\Omega$ and $y_\Omega$ are the full transverse components to linear order. The initial transverse state is specified by $\delta s_{\Omega,-}(0)$. Equation~\eqref{eq:bec-linear} is linear and can be solved exactly. We use a one-sided Fourier transform with complex frequency $z=\omega+i\eta_z$, where $\eta_z>0$ ensures convergence. Throughout this subsection, a tilde denotes the corresponding transformed quantity:
\begin{equation}
\widetilde{\delta s}_{\Omega,-}(z)
=\int_0^\infty dt\,e^{izt}\delta s_{\Omega,-}(t),
\qquad
\widetilde{\delta\Delta}(z)
=\int_0^\infty dt\,e^{izt}\delta\Delta(t).
\label{eq:bec-transform}
\end{equation}
The physical frequency-domain functions are obtained from the boundary value $\eta_z\to0^+$. Transforming Eq.~\eqref{eq:bec-linear} gives
\begin{equation}
\left(z-2\epsilon_\Omega\right)
\widetilde{\delta s}_{\Omega,-}(z)
=i\delta s_{\Omega,-}(0)-\widetilde{\delta\Delta}(z).
\label{eq:bec-transformed-spin}
\end{equation}
Solving Eq.~\eqref{eq:bec-transformed-spin} for $\widetilde{\delta s}_{\Omega,-}(z)$ and substituting it into the transformed self-consistency relation gives $\widetilde{\delta\Delta}(z)=i\mathcal F_0(z)-\Pi(z)\widetilde{\delta\Delta}(z)$. Therefore,
\begin{equation}
\widetilde{\delta\Delta}(z)
=i\frac{\mathcal F_0(z)}{1+\Pi(z)},
\label{eq:bec-transformed-gap}
\end{equation}
where
\begin{equation}
\mathcal F_0(z)
\equiv\chi N\int d\Omega\,g(\Omega)
\frac{\delta s_{\Omega,-}(0)}{z-2\epsilon_\Omega},
\qquad
\Pi(z)
\equiv\chi N\int d\Omega\,g(\Omega)
\frac{1}{z-2\epsilon_\Omega}.
\label{eq:bec-response-functions}
\end{equation}
The exact collective solution within the polarized approximation is
\begin{equation}
\delta\Delta(t)
=i\int_{\mathbb R+i0^+}\frac{dz}{2\pi}\,
e^{-izt}\frac{\mathcal F_0(z)}{1+\Pi(z)}.
\label{eq:bec-exact-gap}
\end{equation}

The long-time behavior follows from deforming the integration contour into the lower half of the complex $z$ plane. For the Gaussian distribution $g(\Omega)$, $\Pi(z)$ can be analytically continued from $\operatorname{Im}z>0$ through the real axis; we denote this continuation by $\Pi^{\mathrm{II}}(z)$. Let the zero of the denominator closest to the real axis be
\begin{equation}
1+\Pi^{\mathrm{II}}(z_{\rm p})=0,
\qquad
z_{\rm p}=\omega_{\rm BEC}-i\gamma_{\rm BEC},
\qquad \gamma_{\rm BEC}>0.
\label{eq:bec-pole}
\end{equation}
Provided that $\mathcal F_0(z_{\rm p})\neq0$ and that no other singular contribution decays more slowly, the residue of this pole gives the leading collective response,
\begin{equation}
\delta\Delta(t)
\approx \mathcal R_{\rm BEC}e^{-iz_{\rm p}t}
=\mathcal R_{\rm BEC}
e^{-\gamma_{\rm BEC}t}e^{-i\omega_{\rm BEC}t},
\label{eq:bec-exponential-gap}
\end{equation}
where $\mathcal R_{\rm BEC}$ is fixed by the initial transverse texture. Thus, under these conditions, the Higgs-oscillation envelope in the polarized BEC-like limit decays exponentially at the rate $\gamma_{\rm BEC}$.

Experimentally, however,  it is very hard to observe the predicted exponential decay of the Higgs oscillations in the BEC regime.
 Since the oscillations damp rapidly,  their intrinsic decay cannot be reliably distinguished from the additional exponential damping due to light scattering. The fitted $\alpha$  we report in the main text therefore describes the effective decay within the experimentally accessible time window, rather than the intrinsic asymptotic decay law. We explain  and justify the fitting procedure in the section below. 

\subsection{Fitting and analysis of the Higgs oscillations}

\begin{figure}
\includegraphics[width=18cm]{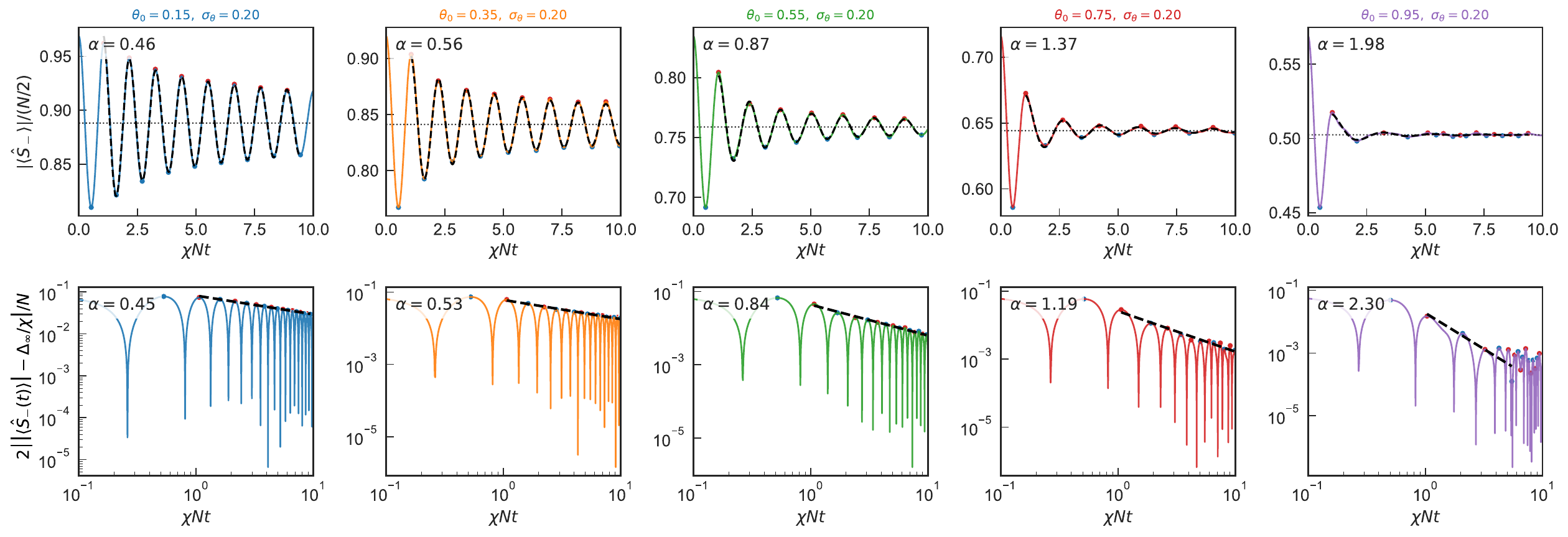}
\caption{Fitting examples for initial tipping angles $\theta_0=0.15,0.35,0.55,0.75$, and $0.95$, with an rms angular spread $\sigma_\theta=0.2$ and $N=20000$ atoms. The red (blue) dots denote the maxima (minima) of the Higgs oscillations. The black curves show the fitting results. Top row: direct fits of $|\langle\hat{S}_-\rangle|/(N/2)$ using a power-law decaying oscillatory function. Bottom row: log-log fits of the oscillation amplitude, obtained from the absolute residual $2/N\bigl||\langle\hat{S}_-\rangle|-\Delta_{\infty}/\chi\bigr|$, to extract the decay exponent.
}\label{fig:s3}
\end{figure}

As a complement to the analytical results in Sec.~\ref{sec:approx}, we extract the Higgs-oscillation decay exponent numerically using two complementary fitting procedures.
First, we fit the full collective-spin dynamics to
\begin{equation}
\chi \frac{|\langle \hat S_-(t)\rangle|}{N/2}=\frac{\Delta_\infty}{N/2}(1+\eta\frac{\cos(\omega_H t+\phi)}{t^\alpha}), \label{eq:supp_order_param}
\end{equation}
where $\Delta_\infty$, $\eta$, $\alpha$, $\omega_H$, and $\phi$ are fitting parameters. The upper row of Fig.~\ref{fig:s3} shows these time-domain fits.

As an independent check, we subtract the asymptotic value and analyze the extrema of the absolute residual, $\bigl||\langle\hat S_-(t)\rangle|-\Delta_\infty/\chi|$. For a power-law envelope, these extrema scale as $t^{-\alpha}$ and therefore lie on a straight line in a log-log plot. We fit this envelope in the log-log representation, as shown in the lower row of Fig.~\ref{fig:s3}.

The close agreement between the two fitting procedures indicates that the extracted decay exponent is robust with respect to the fitting method within the numerical time window considered. In the BEC regime, where the late-time oscillations are too weak to resolve reliably, the fitted $\alpha$ characterizes the decay over this finite accessible time window and should not be interpreted as a determination of the asymptotic exponent.

\subsection{Approximate solution for the Lax roots \label{sec:lax}}
\begin{figure}
\includegraphics[width=10cm]{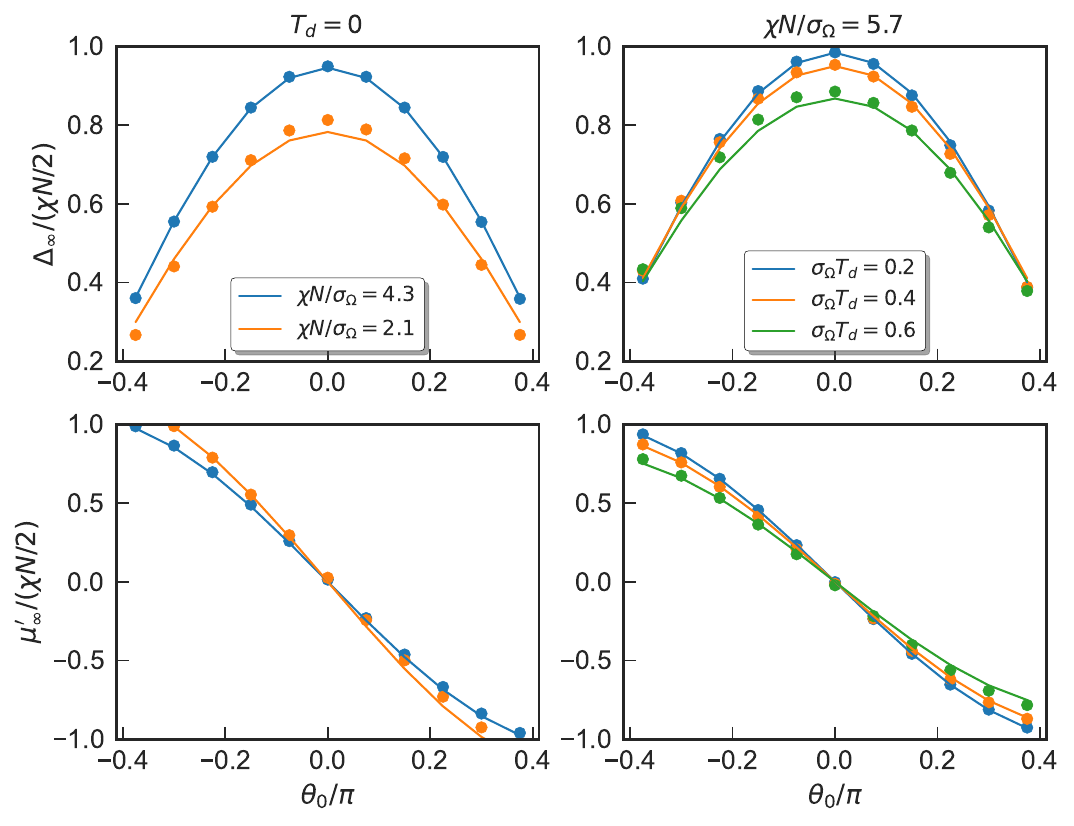}
\caption{Comparison between the approximate Lax solution (solid lines) and
the exact numerical results (dots). In the left panel, we set the
dephasing time $ T_d=0$ and observe that the approximate solution
agrees well with the exact result in the regime $\sigma_\Omega/(\chi N)\ll1$.
In the right panel, we fix $\sigma_\Omega/\chi N$ while varying
$ T_d$. The approximation remains accurate in the regime $\sigma_\Omega T_d \ll 1$.}\label{fig:s2}
\end{figure}
Having introduced the importance of $\Delta_{\infty}$ and $\mu_{\infty}$ for the long-time dynamics, we now derive approximate expressions for them using the Lax analysis~\cite{yuzbashyan2015quantum}. The Lax vector is defined by
\begin{equation}
L^x(u) = \sum_j \frac{s_{j,x}^{(0)}}{u-\Delta\Omega_j} \quad L^y(u) = \sum_j \frac{s_{j,y}^{(0)}}{u-\Delta\Omega_j} \quad  L^z(u) = \sum_j \frac{s_{j,z}^{(0)}}{u-\Delta\Omega_j} - \frac{1}{\chi},
\end{equation}
which depends only on the initial spin configuration, the shifted dispersions $\{\Delta\Omega_j\}$, and the interaction strength $\chi$. The complex roots $u_{n}=\mu_\infty+ n i\Delta_\infty$ with $n=\pm 1$ are determined by $\vec L(u)\cdot\vec L(u)=0$ and encode the long-time asymptotic parameters.

 Substituting the initial condition Eq.~\eqref{eq:fgs}, the Lax equation in the continuum limit becomes:
\begin{equation}
\left(\int d\Omega g(\Omega)\frac{\cos(\theta_{0}+\delta\Omega T_d)}{2u-2\Omega}\right)^{2}+\left(\int d\Omega g(\Omega)\frac{\sin(\theta_{0}+\delta\Omega T_d)}{2u-2\Omega}+\frac{1}{\chi N}\right)^{2}=0,
\end{equation}
where the sum over $j$ has been replaced by the continuum integral over $\Omega$. This expression can be written in a more compact form as
\begin{equation}
\int d\Omega g(\Omega)\frac{e^{-in(\theta_{0}+\delta\Omega T_d)}}{2\Omega-2u_n}=\frac{in}{\chi N}.
\end{equation}
Its imaginary and real parts reproduce Eqs.~\eqref{eq:imag}, respectively.

Using the centered moments of $g(\Omega)$, we expand the phase factor to second order in $\delta\Omega T_d$ in the regime $\sigma_\Omega T_d\ll1$ and obtain
\begin{equation}
\int d\Omega\frac{g(\Omega)}{2\Omega-2u_n}
\approx
\frac{in\left(\frac{e^{-in\theta_0}}{\chi N}-T_d\right)+(u_n+\tilde\mu)T_d^2}
{1+2in(u_n+\tilde\mu)T_d-2(u_n+\tilde\mu)^2T_d^2}.
\end{equation}
Terms beyond second order in $\sigma_\Omega T_d$ have been omitted. Because $g(\Omega)$ is centered at $-\tilde\mu$, the same integral can be expanded for $|u_n+\tilde\mu|\sim\chi N\gg\sigma_\Omega$ as
\begin{equation}
\int d\Omega\frac{g(\Omega)}{2u_n-2\Omega}
\approx
-\frac{1}{2(u_n+\tilde\mu)}
-\frac{\sigma_\Omega^2}{[2(u_n+\tilde\mu)]^3}.
\end{equation}
For $\sigma_\Omega=0$, the root satisfies
\begin{equation}
u_n+\tilde\mu=\frac{in\chi N}{2}e^{in\theta_0}.
\end{equation}
Separating this root into its real and imaginary parts gives
\begin{equation}
\mu'_\infty=-\frac{\chi N}{2}\sin\theta_0,
\qquad
\Delta_\infty=\frac{\chi N}{2}\cos\theta_0.
\end{equation}
For finite $\sigma_\Omega$, solving the root equation perturbatively to leading order in $\sigma_\Omega^2$ gives
\begin{equation}
u_n=-\tilde\mu+\frac{in\chi N}{2}e^{in\theta_0}
-\frac{in}{2}\sigma_\Omega^2\left(\frac{e^{-in\theta_0}}{\chi N}-T_d+\frac{\chi N T_d^2}{2e^{-in\theta_0}}\right).
\end{equation}
Separating the real and imaginary parts gives
\begin{equation}
\begin{aligned}
\mu'_{\infty} &=-\frac{\chi N}{2}\sin\theta_{0}\left(1+\frac{\sigma_{\Omega}^{2}}{(\chi N)^{2}}-\frac{\sigma_{\Omega}^{2}T_d^{2}}{2}\right), \\
\Delta_{\infty} & =\frac{\chi N}{2}\cos\theta_{0}\left(1-\frac{\sigma_{\Omega}^{2} T_d^{2}}{2}-\frac{\sigma_{\Omega}^{2}}{(\chi N)^{2}}\right)+\frac{\sigma_{\Omega}^{2}}{2} T_d,
\end{aligned}
\end{equation}
These expressions give the leading corrections from the inhomogeneity and the initial dephasing. Fig.~\ref{fig:s2} compares them with exact numerical solutions of the Lax equation. The agreement is good within the perturbative regimes assumed in the derivation: $\sigma_\Omega/(\chi N)\ll1$ and $\sigma_\Omega T_d\ll1$.

\section{Effect of finite radial temperature on the Higgs dynamics}
\begin{figure}
    \centering
    \includegraphics[width=0.9\linewidth]{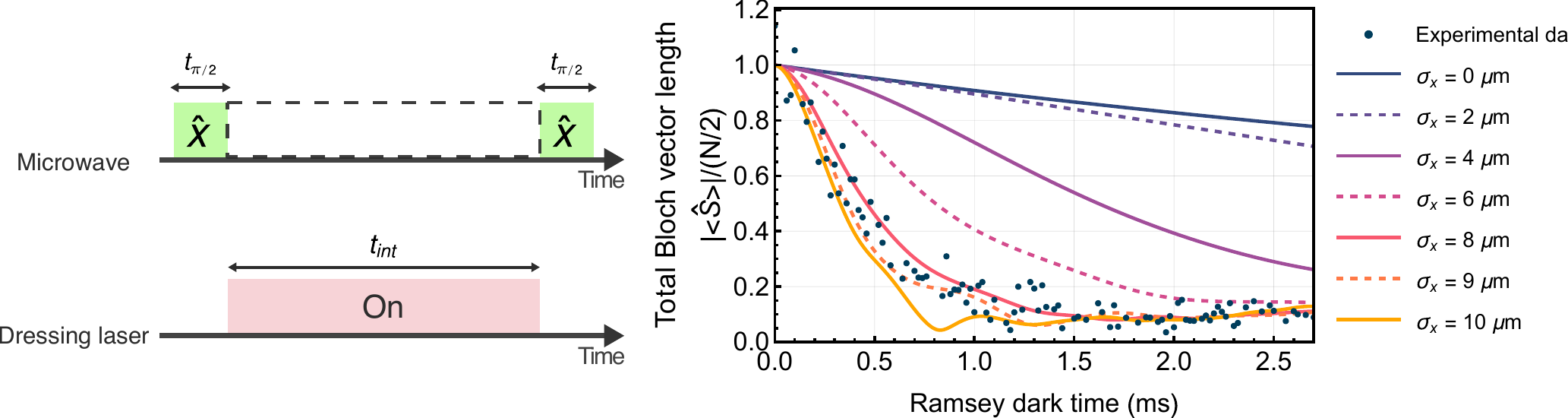}
    \caption{\textbf{Ramsey contrast decay in the presence of the dressing laser.}
The Ramsey sequence is performed between the clock states
$|\Uparrow\rangle \leftrightarrow |\Downarrow\rangle$, for which dephasing due to
magnetic-field inhomogeneity is strongly suppressed. Spatial variation of the
dressing-light intensity, together with the finite radial temperature of the atoms,
produces an inhomogeneous differential AC Stark shift that leads to a shrinking of the collective Bloch vector length.  The left panel shows the experimental sequence, with the dressing laser
applied during the Ramsey dark time $t_{\mathrm{int}}$. The measurement of the length of the Bloch vector after the second $\pi/2$ pulse is achieved using the measurement sequence described in the following section. The right panel shows the
measured total Bloch vector length (blue points) and numerical simulations (solid curves) for
rms transverse spreads $\sigma_x = \sigma_y= 0,\,2,\,4,\,6,\,8,\,9,$ and
$10~\mu\mathrm{m}$.}
    \label{fig:ramsey_decay_with_pump}
\end{figure}

The homogeneous-coupling simulations discussed above yield a fitted Higgs-decay exponent of $\alpha\simeq0.4$ in the BCS-like regime, below the experimental value $\alpha=0.65(8)$. We now investigate whether the coupling inhomogeneity associated with the finite radial temperature of the cloud can account for this difference.

Although rapid axial motion averages over the standing-wave structure of the cavity field, the atoms remain confined in the transverse plane by the combined potentials of the blue-detuned Laguerre--Gaussian beam and the Gaussian cavity mode. Near the atomic cloud, the resulting potential is well approximated by
\begin{equation}
V_\perp(x,y)\simeq \frac{1}{2}m\omega_r^2(x^2+y^2),
\end{equation}
with $\omega_r=2\pi\times700~\mathrm{Hz}$. The transverse motional states can therefore be described by the harmonic-oscillator states $\ket{n_x,n_y}$. At finite temperature, these states are thermally populated, with occupation probability along either transverse direction
\begin{equation}
P(n)=\frac{1}{1+\bar n}
\left(\frac{\bar n}{1+\bar n}\right)^n.
\end{equation}
The mean occupation $\bar n$ determines the rms cloud width through
\begin{equation}
\sigma_r^2=a_r^2\left(\bar n+\frac{1}{2}\right),
\qquad
a_r=\sqrt{\frac{\hbar}{m\omega_r}},
\end{equation}
where $\sigma_r=\sigma_x=\sigma_y$ is the rms size along either transverse direction.  As described below, we estimate $\sigma_x=\sigma_y=8.5~\mu\mathrm{m}$ or a mean thermal occupation $\bar n_x=\bar n_y\simeq 435$.

This motional distribution translates into a distribution of atom--cavity couplings because different harmonic-oscillator states sample different average cavity intensities. The single-photon atom-cavity coupling follows the transverse cavity-field profile, $g(r)=g_0e^{-r^2/w_0^2}$ with $r^2=x^2+y^2$ and $w_0=72~\mu\mathrm{m}$, and $g_0$ denotes the peak atom-cavity coupling strength at the center. 
The effective dispersive coupling of an atom $i$ in state $\ket{n_x,n_y}$ is determined by
\begin{equation}
\frac{g_{\mathrm{eff},i}(n_x,n_y)}{g_{\mathrm{eff}}}
=
\bra{n_x,n_y}
e^{-2(x^2+y^2)/w_0^2}
\ket{n_x,n_y}.
\end{equation}
For $\sigma_r\ll w_0$, expanding the Gaussian profile gives $g_{\mathrm{eff},i}/g_{\mathrm{eff}}\simeq 1-2a_r^2(n_{x,i}+n_{y,i}+1)/w_0^2$. Since $n_x$ and $n_y$ are independently thermally distributed with variance $\mathrm{Var}(n)=\bar n(\bar n+1)$, the rms atom-to-atom coupling variation is approximately $\delta g_{\mathrm{eff}}/\bar g_{\mathrm{eff}}\simeq 2\sqrt{2}\sigma_r^2/w_0^2\sim 0.04$. Using the notation of the main text, $\chi_{ij}=\chi(1+\zeta_{ij})$, the full thermal harmonic-oscillator distribution at $\sigma_r=8.5~\mu\mathrm{m}$ gives $\chi(\sigma_r=8.5~\mu\mathrm{m})/\chi(\sigma_r=0)\simeq 0.90$ and standard deviation $\mathrm{std}(\zeta_{ij})\simeq 0.053\ll1$.

We incorporate this variation into numerical simulations by independently sampling $n_x$ and $n_y$ for each atom from the thermal distribution and solving the atom--cavity mean-field equations in Eq.~\eqref{eq:ac} with the corresponding atom-dependent couplings. The sampled motional states remain fixed throughout the evolution, neglecting mode-changing collisions and cavity-induced transitions between radial levels.

To determine the mean occupation entering these simulations, we measured the Bloch vector length after the Ramsey sequence shown in Fig.~\ref{fig:ramsey_decay_with_pump} left panel.  After the final $\pi/2$ pulse, the Bloch vector length is measured as described in section \ref{sec:measurement}.
The dressing laser is applied during the Ramsey dark time.  Its intensity varies radially, producing differential AC-Stark shifts between $\ket{\Uparrow}$ and $\ket{\Downarrow}$ that depends on the motional quantum states $n_x$ and $n_y$. The resulting phase dispersion between different atoms in different motional states reduces the length of the total Bloch vector as shown in Fig.~\ref{fig:ramsey_decay_with_pump} right panel. Because the sequence probes the clock states, whose differential magnetic sensitivity is strongly suppressed, dephasing from magnetic-field inhomogeneity is negligible. We then numerically simulate the measured Bloch vector length versus time for different $\bar{n}_x=\bar{n}_y$ or equivalently $\sigma_x=\sigma_y$. From comparing the simulations and the data, we estimate that $\sigma_x=8.5(5)~\mu\mathrm{m}$.


We use the Ramsey-inferred rms extent $\sigma_x=8.5~\mu\mathrm{m}$ in the collective-dynamics simulations to produce the dashed grey curve of Fig.~4d in the main text, with solid grey curves corresponding to $\sigma_x=0$.  In the BCS limit, the numerical simulation with $\sigma_x=8.5~\mu\mathrm{m}$ produces a fitted Higgs-decay exponent of $\alpha=0.58$, consistent with the experimental value $\alpha=0.65(8)$ within its uncertainty. Thus, the small radial coupling inhomogeneities $\zeta_{ij}$ inferred from Ramsey spectroscopy accounts for the additional damping of the Higgs oscillations relative to the homogeneous-coupling  simulations where $\zeta_{ij}=0$.

\section{Non-destructive readout of the total Bloch vector length}
\label{sec:measurement}
\begin{figure}
    \centering
    \includegraphics[width=0.8\linewidth]{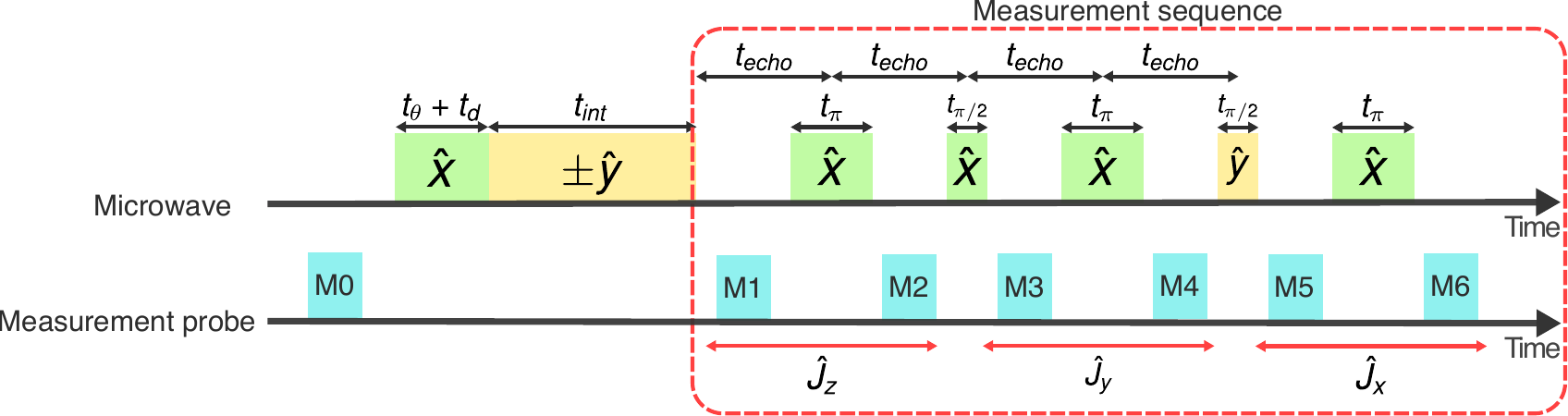}
    \caption{Experimental pulse sequence depicted in the bare-spin basis. The pulse sequence inside the dashed box represents the state-readout sequence.}
    \label{fig:meas_seq}
\end{figure}
To measure the total collective vector length in a single experimental shot, we perform non-destructive measurements of all three collective spin components (Fig.~\ref{fig:meas_seq}). Each component is read out via a quantum non-demolition (QND) measurement of the population difference. The QND measurement extracts the time-averaged cavity resonance frequency shift, which depends approximately linearly on the populations in $\ket{\Uparrow}$ and $\ket{\Downarrow}$:
\begin{align}
    \Delta f = \chi_{\text{QND}\Uparrow}N_{\Uparrow} + \chi_{\text{QND}\Downarrow}N_{\Downarrow}.
\end{align}
This is implemented by sweeping a probe beam across the cavity resonance and detecting the reflected light via homodyne detection. For each spin component, the cavity is swept twice with a $\pi$ pulse in between. The population difference is extracted from the difference of the two frequency shift measurements:
\begin{align}
    \Delta f_{M1} - \Delta f_{M2} = (\chi_{\text{QND}\Uparrow} - \chi_{\text{QND}\Downarrow})(N_\Uparrow - N_\Downarrow)
\end{align}
where $\chi_{\mathrm{QND}\Uparrow}= 2\pi \times 12.69~\text{Hz}$ and $\chi_{\mathrm{QND}\Downarrow}=2\pi \times 6.15 ~\text{Hz}$ for an atom-cavity detuning magnitude $|\Delta_a|=2\pi \times 6.289~\text{GHz}$ from the cycling transition.
The probe power is kept low, with a measurement strength 9.5~dB above the standard quantum limit for $N = 120{,}000$, ensuring that measurement back-action does not affect subsequent measurements.

After the interaction, we first measure the $J_z$ component directly via QND. We then apply a $\frac{\pi}{2}$ pulse to rotate $J_y$ onto the population difference for readout. To preserve the contrast, we use a spin-echo-like sequence to remove phase shifts imposed by the probe light.  Specifically, the measurement pulse timing is chosen such that the interval between when the microwave dressing is turned off and the center of the $\pi$ pulse between the first and second dispersive measurements ($t_{\text{echo}}$) equals the interval between that $\pi$ pulse and the subsequent $\frac{\pi}{2}$ pulse. The $J_x$ component is measured using the same sequence, with the phase of the $\frac{\pi}{2}$ rotation shifted by $\frac{\pi}{2}$ radians to project $J_x$ onto the population difference.

Additionally, we measure the total atom number at the beginning of every experimental sequence after optically pumping the atoms into $\ket{\Uparrow}$. The measured atom number is typically $N=120{,}000$, which we use as the reference atom number when characterizing the probe strength relative to the standard quantum limit. This number is larger than the $N=87{,}500$ atoms participating in the coherent dynamics discussed in the main text, primarily due to atom loss from photon scattering during the initial atom-number measurement and imperfect optical pumping.
\section{OAT Interaction Strength: Dressed vs. Undressed Basis}
\begin{figure}[htb]
    \centering
    \includegraphics[width=0.7\linewidth]{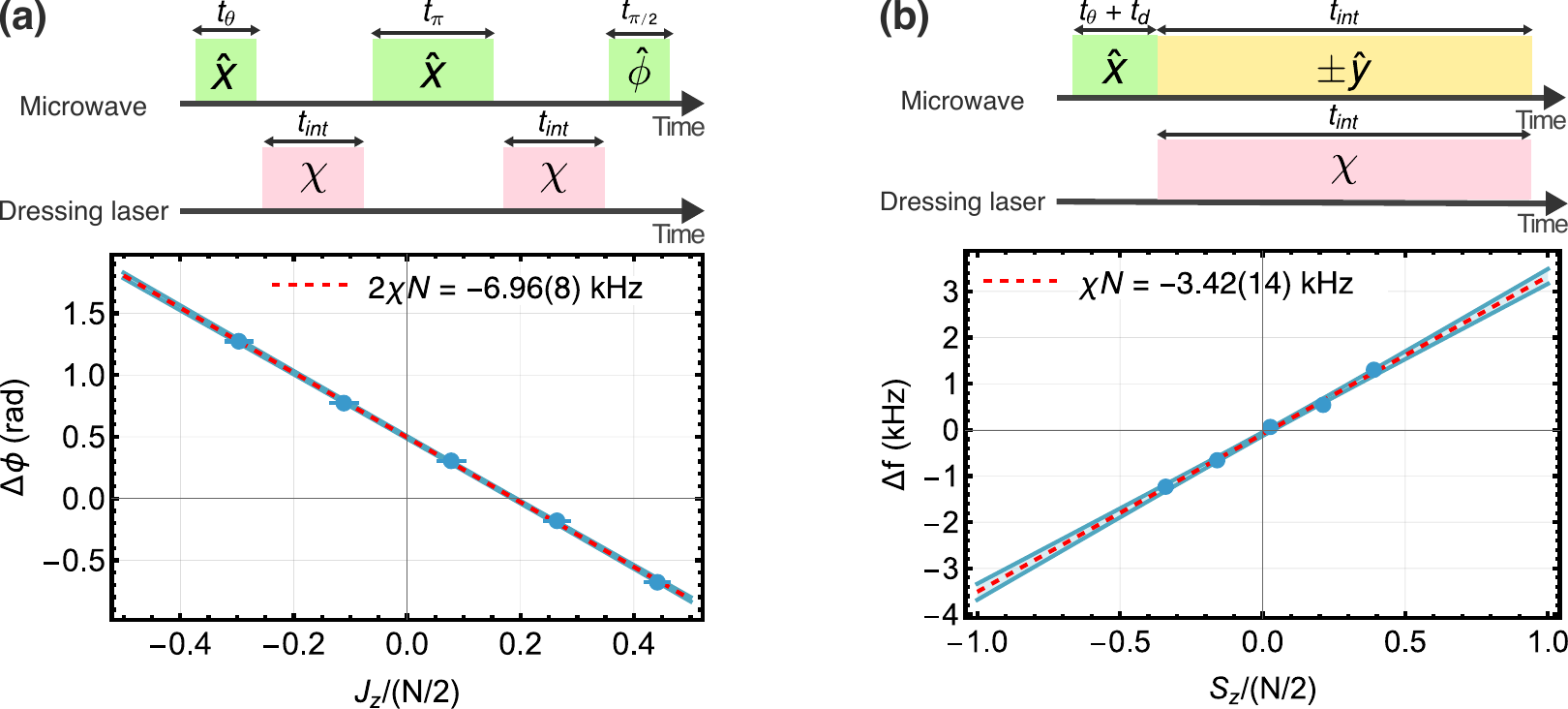}
    \caption{\textbf{Comparison of Ising (OAT) dynamics with and without microwave dressing.} (a) Cavity-feedback Ising dynamics, $\hat H_{\rm Ising}/\hbar=-2\chi\hat J_z^2$, in the bare-spin basis. (b) The exchange interaction, $\hat H_{\rm ex}/\hbar=-\chi(\hat S_x^2+\hat S_y^2)\approx -\chi\left(\mathbf{\hat{S}}^2- \hat{S}_z^2 \right)$, in the dressed basis. Thus, the microwave Floquet dressing reduces the coefficient of the two Ising interactions $\hat{J}_z^2 \rightarrow \hat{S}_z^2$ by a factor of two. The top panels show the corresponding pulse sequences used to measure the dynamics in each case. For the sequence in (a), an interaction time of $t_{\text{int}} = 30~\mu\text{s}$ (total interaction time $2t_{\text{int}} = 60~\mu\text{s}$) was used.}
    \label{fig:OAT_comp}
\end{figure}
Without microwave dressing, our system realizes the Ising, or one-axis-twisting (OAT), Hamiltonian $\hat{H}_{\rm Ising}/\hbar = -2\chi \hat{J}_z^2$ via cavity-feedback dynamics~\cite{leroux2010implementation,schleiter-smith_squeezing_2010,Thompson2022SAI}. This Hamiltonian is generated by applying a dressing laser at fixed detuning $\Delta_d$. The spin population tunes the dressed cavity resonance, causing the intracavity photon number $\hat{a}^\dagger\hat{a}$ to depend linearly on the population difference $\hat{J}_z$, which maps the dispersive coupling term $\hat{H}_{\rm disp}\propto \hat{a}^\dagger\hat{a}\hat{J}_z$ onto the Ising Hamiltonian.

In the presence of on-resonant microwave dressing, the time-averaged Hamiltonian, neglecting single-particle terms, is
\begin{align}
    \hat{H}_{\rm avg}/\hbar = -\chi\left(\hat{J}_z^2 + \hat{J}_y^2\right) = -\chi\left(\hat{\mathbf J}^2 - \hat{J}_x^2\right),
\end{align}
which in the dressed basis takes the form
\begin{align}
    \hat{H}_{\rm ex}/\hbar = -\chi\left(\hat S_x^2+\hat S_y^2\right)=-\chi\left(\hat{\mathbf S}^2 - \hat{S}_z^2\right).
\end{align}
The factor of $1/2$ relative to the bare-basis Ising interaction arises from time averaging under the continuous microwave drive.

We observe interaction-induced dynamics both without (Fig.~\ref{fig:OAT_comp}a) and with (Fig.~\ref{fig:OAT_comp}b) microwave dressing and estimate the collective interaction strength in each case. Without microwave dressing, the Ising interaction-induces an accumulated precession phase shift about $\hat{z}$ at expected frequency $2\chi N$.  We look for this induced phase shift  using a spin-echo sequence (top panel of Fig.~\ref{fig:OAT_comp}a), in which the interaction is applied during the dark time and the induced phase shift is measured as a function of the initial population difference $\hat{J}_z$. With microwave dressing, the exchange interaction's Ising term induces a mean-field precession about  the microwave-drive axis at expected frequency $\chi N$. This appears as a  shift of the measured Rabi flopping frequency as a function of the population difference $\hat{S}_z$ in the dressed basis (top panel of Fig.~\ref{fig:OAT_comp}b). Both measurements were performed with atom number $N \approx 87{,}500$, intracavity photon number $|\beta|^2 \approx 3200$, and mean detuning $\Delta_d = -2\pi\times1.9~\text{MHz}$.

In both cases, the observed interferometric phase shift (or apparent Rabi-frequency shift) scales linearly with the initial projection $J_z$ (or $S_z$). The two fitted slopes have opposite signs under the phase conventions of the corresponding measurement sequences. Quoting positive interaction-strength magnitudes, we obtain $2\chi N/(2\pi) = 6.96(8)~\text{kHz}$ without microwave dressing and $\chi N/(2\pi) = 3.42(14)~\text{kHz}$ with microwave dressing. The ratio of the latter to the former is $0.49(2)$, consistent with the expected ratio of $1/2$.

    \label{fig:ramsey_gap}

\end{document}